\documentclass[onecolumn]{raa}
\usepackage{natbib, graphicx, amsmath,gensymb}
\bibpunct{(}{)}{;}{a}{}{,}

\usepackage{hyperref}
\hypersetup{pdftitle = The title of my PDF, pdfauthor = My name, pdfsubject= The subject, pdfkeywords = keyword1 keyword2 keyword3} 
\hypersetup{colorlinks = true, linkcolor = green, anchorcolor = red, citecolor = blue, filecolor = red, urlcolor = red}

\include{aas_journals}

\def\rmd{{\mathrm d}}
\newcommand{\vect}[1]{\mathbf{#1}}
\newcommand{\msun}{h^{-1} M_{\sun}}
\newcommand{\avg}[1]{\langle{#1}\rangle}
\newcommand{\abs}[1]{\left\vert{#1}\right\vert}

\makeatletter

\newcommand{\Rmnum}[1]{\expandafter\@slowromancap\romannumeral #1@}
\makeatother
\begin{document}

\title{Estimating power spectrum of discrete cosmic momentum field with fast Fourier transform}

\volnopage{ {\bf 20XX} Vol.\ {\bf X} No. {\bf XX}, 000--000}
   \setcounter{page}{1}

\author{Jun Pan \inst{1}}
\institute{
National Astronomical Observatories, Chinese Academy of Sciences, 20A Datun Rd., 
Beijing 100101, P. R. China {\it jpan@bao.ac.cn}\\
\vs \no
   {\small Received~~20xx month day; accepted~~20xx~~month day}
}

\abstract{
Fast Fourier transform based estimators are
formulated for measuring momentum power spectra, including the auto power spectra of the momentum, 
the momentum divergence, and the cross spectrum of density fluctuation and momentum divergence.
Algorithms using the third order Bettle-Lemari\'{e} scaling function to assign discrete objects to regular grids for
fast Fourier transform are proposed to clean alias effects. Numerical experiments prove that the implementation
can achieve sub-percent precision till close to the Nyquist frequency. 
Impact of removing bulk flow on estimation of momentum power spectra are derived theoretically and verified
numerically, subtracting bulk flow has little effects at large scales
but might induce meaningful differences in nonlinear regime, and probably it is not necessary to 
subtract bulk flow for samples which peculiar velocities are exact or sufficiently accurate. 
Momentum power spectra of dark matter samples from N-body simulation are measured and discussed.
As expected, prediction of the one loop Eulerian perturbation theory
agrees with simulation only slightly better than the linear theory at $z=0$, but can be applied to
higher redshift with improved accuracy.
Measurements of simulation data and the one loop Eulerian theory both
reveal that the momentum field contains strong rotational part, and there is a large stochastic component in the
divergence of momentum which is not correlated with the density field. The three kinds of momentum
power spectra have their own characteristics.
}

\authorrunning{J. Pan }            
   \titlerunning{Estimating power spectrum of momentum}  

\maketitle

%%======================================================
\section{Introduction}
The cosmic momentum, as product of the dimensionless density
and the peculiar velocity, is essentially the core of  
velocity correlation functions \citep[e.g.][]{GorskiEtal1989, WangEtal2018} and 
the kinematic Sunyaev-Zel'dovich effect \citep[kSZ, e.g.][]{MaFry2002, ParkEtal2016}. 
There is also strong link between the divergence of cosmic momentum and the
Rees-Sciama effect \citep[e.g.][]{Seljak1996} and the integrated Sachs-Wolfe 
effect \citep[ISW, e.g.][]{SmithEtal2009}. Much attention have been paid to realize the 
potential of momentum power spectrum in cosmology, 
including attempt to develop theoretical 
models \citep[e.g][]{OkumuraEtal2014, CarrascoEtal2014, 
SenatoreZaldarriaga2015, SugiyamaEtal2016} and 
practices of probing the physical Universe \citep{Park2000, ParkPark2006, QinEtal2019}.

To facilitate research on cosmic momentum, reliable and accurate 
algorithms to estimate power spectrum of cosmic momentum is pivotal. 
One of the benefit of working with momentum is that there is not such 
annoying uneven sampling problem
as in the analysis of the volume-weighted peculiar velocity field.
If it is the volume-weighted velocity field to be explored, special algorithms have 
to be devised to resample the peculiar velocity field, such as algorithms implemented with Delaunay or Voronoi tessellation \citep[e.g.][]{BernardeauWeygaert1996, PueblasScoccimarro2009}, 
or interpolation based on various kernel 
functions \citep[e.g.][]{ColombiEtal2007, ZhengEtal2013, YuEtal2015}. 
Even armed with these tools, accuracy control is yet very challenging to the
estimation of statistics of volume-limited velocity fields, which actually varies by cases. 
In contrast, the algorithm of measuring momentum spectrum effectively is similar to that of the
density power spectrum, as already shown by \citet{Park2000}, \citet{ParkPark2006} and 
\citet{Howlett2019}. In these works, estimators for momentum power spectrum accounting 
for shot noises and proper weights
are proposed and tested, setting up solid basis for relevant applications.
However, if fast Fourier transform (hereafter FFT) is adopted to realize these algorithms, alias
effect could be significant \citep{Jing2005}, which treatment is absent in current procedures.

Meanwhile it is worth of addressing that the momentum and the momentum divergence are different.
The cosmic momentum is mainly related to applications about correlation functions of peculiar velocities, 
while the momentum
divergence is connected to cosmological probes about temporal evolution of 
gravitational potential. Momentum field
is composed of its potential and curl components, mathematically it is quite simple to take the
spatial derivative of the momentum field to generate its divergence, but numerically measuring
power spectrum of momentum divergence would require a different estimator which is not
explicitly presented. Thereof the main purpose of this report 
is to present a formal derivation and description of FFT based estimators 
of the auto power spectra of the cosmic momentum field, momentum divergence and 
the cross spectrum of density and momentum divergence, with appropriate 
prescription for cleaning shot noise and aliasing effect. As it is straightforward
to apply these algorithms to non-uniform samples, estimators presented here are about ideal
samples free of effects of selection functions, geometric masking and etc.

In the next section, we will present algorithms for estimation of
power spectra of momentum. Section 3 is dedicated to investigation on effects of subtracting bulk flow, 
momentum spectra of dark matter samples of a N-body simulation are explored in
Section 4.  The last section is of discussion and conclusion.
%======================================================

%%=================================================================
\section{Estimators}
\subsection{Auto power spectrum of the cosmic momentum}
At a given position $\vect{r}$ , the cosmic momentum of dark matter or structures like halos or galaxies, is defined by
\begin{equation}
\vect{p}(\vect{r})\equiv\left(1+\delta(\vect{r})\right) \vect{v}(\vect{r})
\end{equation}
with $\delta$ being the number density contrast and $\vect{v}$ the peculiar velocity.
For a sample of volume $V_S$, in Fourier space at wave vector $\vect{k}$ the momentum
can be written in analogues to a vector, 
\begin{equation}
\vect{p}(\vect{k})\equiv\frac{1}{V_S}\int \vect{p}(\vect{r})e^{i\vect{k}\cdot \vect{r}} \rmd \vect{r} = 
\frac{1}{V_S} \int \left( {\rm p}_{\hat{e}_1}(\vect{r}),\  {\rm p}_{\hat{e}_2}(\vect{r}),\ {\rm p}_{\hat{e}_3}(\vect{r}) \right)
e^{i\vect{k}\cdot \vect{r}} \rmd \vect{r}
=\left( {\rm p}_{\hat{e}_1}(\vect{k}),\  {\rm p}_{\hat{e}_2}(\vect{k}),\ {\rm p}_{\hat{e}_3}(\vect{k}) \right)\ ,
\end{equation}
the momentum power spectrum is constructed by
\begin{equation}
P_p(\vect{k})=\avg{\vect{p}(\vect{k})\cdot \vect{p}^*(\vect{k})}=\sum_{j=1}^3 
\avg{{\rm{p}}_{\hat{e}_j}(\vect{k}) {\rm{p}}_{\hat{e}_j}^*(\vect{k}) }
\end{equation}
where the supscript $*$ refers to the complex conjugate, $\hat{e}_j$ is one of the three unit 
coordinate vectors defining a three-dimensional Cartesian coordinate system.

Shot noise can be derived following \citet[][\S \ 41]{Peebles1980}. 
The sample space $V_S$ is divided into infinitesimal
cells of volume $\rmd V_{S,j}$ in which number of objects $n_j=1$ or $0$, and 
if $n_j=1$ there is measurement of peculiar velocity $\vect{v}_j$. 
Let $\sum_j n_j=N$ and $\bar{n}=\avg{n_j}=N/V_S$, 
\begin{equation}
\hat{\vect{p}}(\vect{k})= \frac{1}{N}\sum_j n_j \vect{v}_j e^{i \vect{k}\cdot \vect{r}_j}\ ,
\end{equation}
and  
\begin{equation}
\avg{\hat{\vect{p}}(\vect{k}_1) \cdot \hat{\vect{p}}^*(\vect{k}_2)}= \frac{1}{N^2} \sum_{j,\ell,j \neq \ell} 
\avg{ (n_j \vect{v}_j)\cdot (n_\ell \vect{v}_\ell)}  e^{i \vect{k}_1\cdot \vect{r}_j - i \vect{k}_2\cdot \vect{r}_\ell} 
+ \frac{1}{N^2}\sum_j \avg{n_j^2 v_j^2} e^{i (\vect{k}_1-\vect{k}_2) \cdot \vect{r}_j}\ .
\label{eq:calsn}
\end{equation}
Since $n_j=1$ or $0$, $n_j^2=n_j=1$ or $0$, replacing the ensemble average with 
spatial averages yields
\begin{equation}
\begin{aligned}
&\avg{n^2_j v^2_j}=\sum_{j, n_j=1} v^2_j/N\equiv  \hat{\sigma}_v^2 \\
&\avg{(n_j \vect{v}_j)\cdot (n_\ell \vect{v}_\ell)}_{j\neq \ell}={\bar{n}}^2\xi_p(\vect{r}_2 -\vect{r}_1) \rmd V_{S,j} \rmd V_{S,\ell}\ ,
\end{aligned}
\label{eq:rawppnoise}
\end{equation}
where $\xi_p=\avg{\vect{p}(\vect{r}_1)\cdot \vect{p}(\vect{r}_2) }$ is the
scalar two-point correlation function of cosmic momentum.
The raw power spectrum turns to be as simple as 
\begin{equation}
\hat{P}_p(\vect{k})=P_p(\vect{k}) +\frac{\hat{\sigma}_v^2}{N}\ .
\label{eq:psn}
\end{equation}

The aliasing effect is formulated with the approach of \citet{Jing2005}. The sampling function
corresponding to grids for FFT is a sum of Dirac functions
$\Pi(\vect{r}/\Delta L)\equiv \sum_{\vect{J} } \delta_D(\vect{r}/\Delta L-\vect{J})$
in which $\vect{J}$ is an integer vector and $\Delta L$ is the grid spacing. 
Let the window function used to assign objects to grid points be $W$, the raw momentum becomes
$\hat{\vect{p}}(\vect{r})=\Pi \left( \vect{r}/\Delta L \right) \int \vect{p}(\vect{r}_1) W(\vect{r}_1-\vect{r}) \rmd \vect{r}_1 $
so that
\begin{equation}
\hat{\vect{p}}(\vect{k}) = \frac{1}{N} \int \Pi \left( \frac{\vect{r}}{\Delta L} \right) 
\sum_j n_j \vect{v}_j W(\vect{r}_j-\vect{r}) e^{i \vect{k}\cdot \vect{r}}\rmd \vect{r}\ ,
\label{eq:rawFFTed}
\end{equation}
and the power spectrum would be constructed through
\begin{equation}
\begin{aligned}
\langle \hat{\vect{p}}(\vect{k}_1)\cdot    & \hat{\vect{p}}^*(\vect{k}_2) \rangle = 
 \frac{1}{N^2} \int \int   \rmd \vect{r}_1 \rmd \vect{r}_2 
\Pi\left(  \frac{\vect{r}_1}{\Delta L}  \right) 
\Pi \left(  \frac{\vect{r}_2}{\Delta L}  \right) 
e^{i(\vect{k}_1\cdot \vect{r}_1-\vect{k}_2\cdot \vect{r}_2) }  \\
\times & \left[\sum_{j,\ell, j\neq \ell} \avg{(n_j \vect{v}_j) \cdot (n_\ell \vect{v}_\ell)}W(\vect{r}_j-\vect{r}_1)
W(\vect{r}_\ell-\vect{r}_2) + \sum_j \avg{n_j^2\vect{v}_j^2} W(\vect{r}_j-\vect{r}_1) W(\vect{r}_j-\vect{r}_2) \right] 
 \ .
\end{aligned}
\end{equation}
Since
\begin{equation}
\Pi(\vect{k})=\frac{1}{V_s}\int \Pi\left( \frac{\vect{r}}{\Delta L}\right)e^{i \vect{k}\cdot \vect{r}} \rmd \vect{r}=\frac{(2\pi)^3}{V_s}\sum_{\vect{J}}\delta_D(\vect{k}-2k_N \vect{J})
\end{equation}
with $k_N=\pi/\Delta L$ being the Nyquist frequency, there is
the raw momentum power spectrum
\begin{equation}
\hat{P}_p(\vect{k})= P_p(\vect{k})W^2(\vect{k})+\sum_{\vect{J}\neq 0} W^2(\vect{k}+2k_N\vect{J}) P_p(\vect{k}+2k_N \vect{J}) + \frac{\hat{\sigma}_v^2}{N}
\sum_{\vect{J}}W^2(\vect{k}+2k_N\vect{J})\ , 
\label{eq:rawPp}
\end{equation}
which is very similar to the formula of matter power spectrum in \citet{Jing2005} except for a 
factor $\hat{\sigma}_v^2$ topped on the shot noise term, correction methods of \citet{Jing2005}, \citet{CuiEtal2008}, \citet{YangEtal2009} and \citet{ColombiEtal2009}
all can be readily applied.

%%-------------------------------------------------------------------------------------------------------------------------------
\subsection{Auto power spectrum of the momentum divergence}
The momentum divergence $\theta_p\equiv -\nabla \cdot \vect{p}(\vect{r})/(Haf)$ in Fourier space is
$\theta_p(\vect{k})  = i\vect{k}\cdot \vect{p}(\vect{k})/(Haf)$,  
$f=\rmd \log D(a)/ \rmd \log a$ and $D(a)$ is the linear density growth factor at redshift $z=1/a-1$. 
Practically divergence of momentum field is produced through 
$\hat{\theta}_p(\vect{k})=i \vect{k}\cdot \hat{\vect{p}}(\vect{k})/(Haf)$, such that 
\begin{equation}
\hat{P}_{\theta_p}(\vect{k})=\avg{\hat{\theta}_p \hat{\theta}^*_p} = \frac{1}{(Haf)^2} 
\avg{\left[ \vect{k}\cdot \hat{\vect{p}}(\vect{k}) \right] \left[ \vect{k}\cdot \hat{\vect{p}}^*(\vect{k}) \right]}\ .
\end{equation}
Inserting Eq.~\ref{eq:rawFFTed} results in
\begin{equation}
\begin{aligned}
\hat{P}_{\theta_p} (\vect{k}) =P_{\theta_p}(\vect{k})W^2(\vect{k})
+ & \frac{1}{(Haf)^2}\sum_{\vect{J}\neq 0} W^2(\vect{k}+2k_N\vect{J}) 
\avg{ [  \vect{k}\cdot \vect{p}(\vect{k}+2k_N \vect{J}) ] [ \vect{k}\cdot \vect{p}^*(\vect{k}+2k_N \vect{J}) ]  }\\
&+ \frac{k^2}{(Haf)^2} \frac{\avg{v^2\mu^2}}{N}  
\sum_{\vect{J}}W^2(\vect{k}+2k_N\vect{J})\ ,
\end{aligned}
\label{eq:rawPtheta}
\end{equation}
where $\avg{v^2\mu^2}=\sum_j v_j^2 \mu_j^2/N$ with $\mu_j=\vect{k}\cdot \vect{v}_j/(kv_j)$. 

The shot noise in Eq.~\ref{eq:rawPtheta} deserves more attention. For a fair sample, such as
a full simulation data, the condition $\sum_j \vect{v}_j=0$ tells that $\sum v_j\mu_j=0$. By virtue of
isotropy and homogeneity, velocity amplitude $v$ shall not be correlated with its direction $\mu$, 
thus $\avg{v^2\mu^2}=\hat{\sigma}_v^2\avg{\mu^2}$, and $1/3$ could be a convenient approximation 
to $\avg{{\mu}^2}$. 
But for samples constructed from observation of a finite space of the Universe, or 
extracted as subsamples of 
the full simulation, the bulk flow $\vect{v}_b=\sum_j \vect{v}_j/N$ is generally
not zero, the shot noise will be directional dependent. To see the point, let $\vect{v}'=\vect{v}-\vect{v}_b$, 
$\mu'=\vect{k}\cdot \vect{v}'/(kv')$ and $\mu_b= \vect{k}\cdot \vect{v}_b/(kv_b)$, such that
\begin{equation}
\avg{v^2\mu^2}=\avg{{v'}^2{\mu'}^2}+v_b^2\mu_b^2 = \hat{\sigma}_{v'}^2 \avg{{\mu'}^2}+v_b^2\mu_b^2 \ ,
\label{eq:thetanoise}
\end{equation}
in which $\hat{\sigma}_{v'}^2=\sum_{j=1}^N(\vect{v}_j-\vect{v}_b)^2/N$. 
Obviously the shot noise varies with $\mu_b$, 
and the strength of such dependence is determined by amplitude of $v_b$. Of course, 
in isotropic $P_{\theta_p}(k)$, the
$\vect{k}$ directional dependence of shot noise vanishes and
$\avg{v^2\mu^2}=\sigma_{v}^2/3$. 

%%----------------------------------------------------------------------------------------------
\subsection{Cross spectrum of the density and the momentum divergence}
It is fairly trivial to construct the estimator for the cross spectrum, in analogues to last subsection,
\begin{equation}
\begin{aligned}
\hat{P}_{\delta\theta_p}(\vect{k}) =P_{\delta\theta_p}(\vect{k})W^2(\vect{k})-  &\frac{i}{Haf}\sum_{\vect{J}\neq 0} W^2(\vect{k}+2k_N\vect{J}) 
\avg{  \delta(\vect{k}+2k_N \vect{J}) \left[ \vect{k}\cdot \vect{p}^*(\vect{k}+2k_N \vect{J}) \right]  }\\
&- \frac{ik}{Haf} \frac{\avg{v\mu}}{N}  
\sum_{\vect{J}}W^2(\vect{k}+2k_N\vect{J})\ .
\end{aligned}
\label{eq:rawPx}
\end{equation}
It is easy to see that $\avg{v\mu}=v_b\mu_b$, an interesting thing is that non-zero bulk flow induces shot noise in the imaginary part of the cross
spectrum, and such shot noise will be zero in the isotropic power spectrum $P_{\delta\theta_p}(k)$. 

%%----------------------------------------------------------------------------------------------------------------
\subsection{Test with N-body simulation data}
%%-----------------------------------------------------------------------------------------------
\subsubsection{Algorithm setup and data preparation}
FFT is computed with the FFTW3 package\citep{FFTW05}. Assignment of objects to FFT grids 
is implemented with the third-order orthogonalized 
Battle-Lemari\'{e} spline function \citep{YangEtal2009}, practice shows that
adoption of the fifth-order B-spline function brings up minute differences less than $1\%$.

Samples used for our experiments are produced from data sets of a N-body simulation.
The simulation is of pure dark matter and 
realized with the {\it Gadget-2} code \citep{Springel2005}, which assumes a $\Lambda$ cold dark 
matter ($\Lambda$CDM) cosmology model with parameters 
$\Omega_m = 0.26, \; \Omega_b=0.044, \; \Omega_{\Lambda}=0.74, h = 0.71, \; \sigma_8=0.8, \; n_s=1$.
The run consists of $N=1,024^3$ particles within a periodic cubic box of size 
$L_{box}=1000 h^{-1}$Mpc, each particle has mass of $6.72\times 10^{10} \msun$. 
Samples employed in this work include
\begin{enumerate} 
\item outputs of the simulation at picked redshifts, mainly the one at $z=0$ and the initial condition
at $z=80$;
\item ten random samples generated 
from the $z=0$ output by randomly relocating dark matter particles while 
preserving their velocities both in amplitude and
direction.
\item two sets of 64 subsamples at $z=0$ and $z=80$ respectively, constructed by 
evenly splitting the full sample volume into $4\times4\times4$ non-overlapping cubes
of size $250$Mpc/h. 
\end{enumerate} 

%%-----------------------------------------------------------------------------------------------------------------
\subsubsection{Shot noise}

\begin{figure*}
\resizebox{\hsize}{!}{\includegraphics{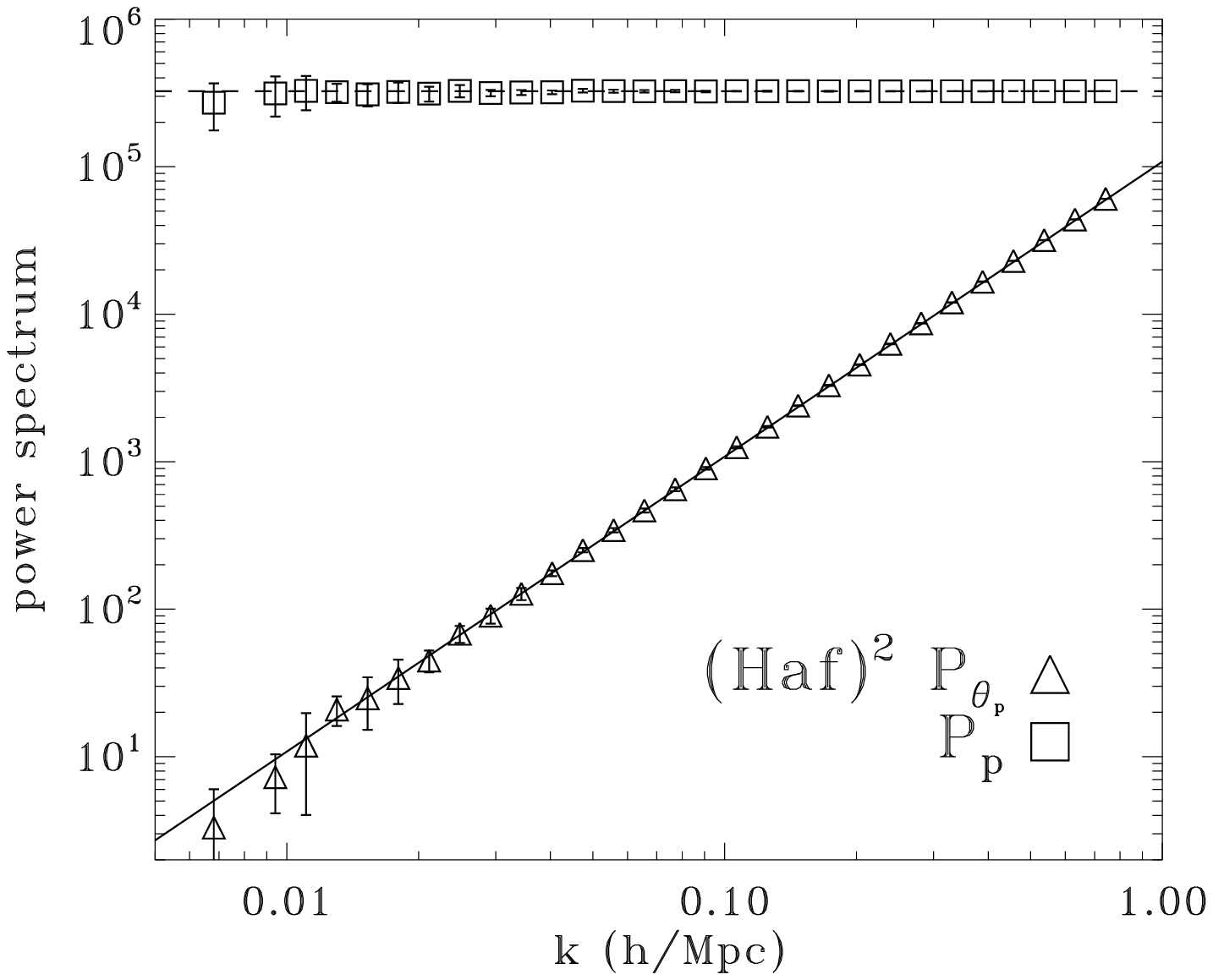}\includegraphics{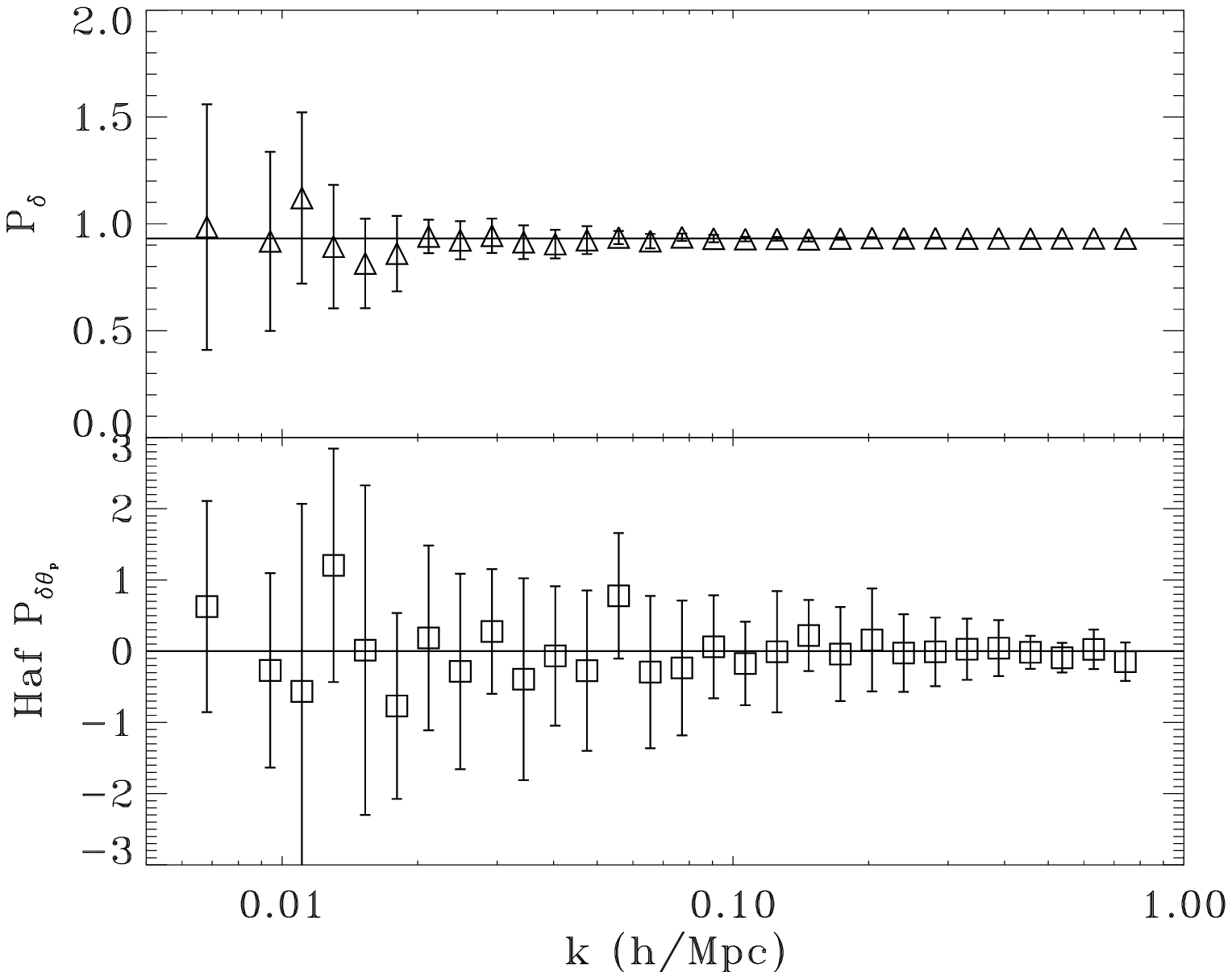}}
\resizebox{\hsize}{!}{\includegraphics{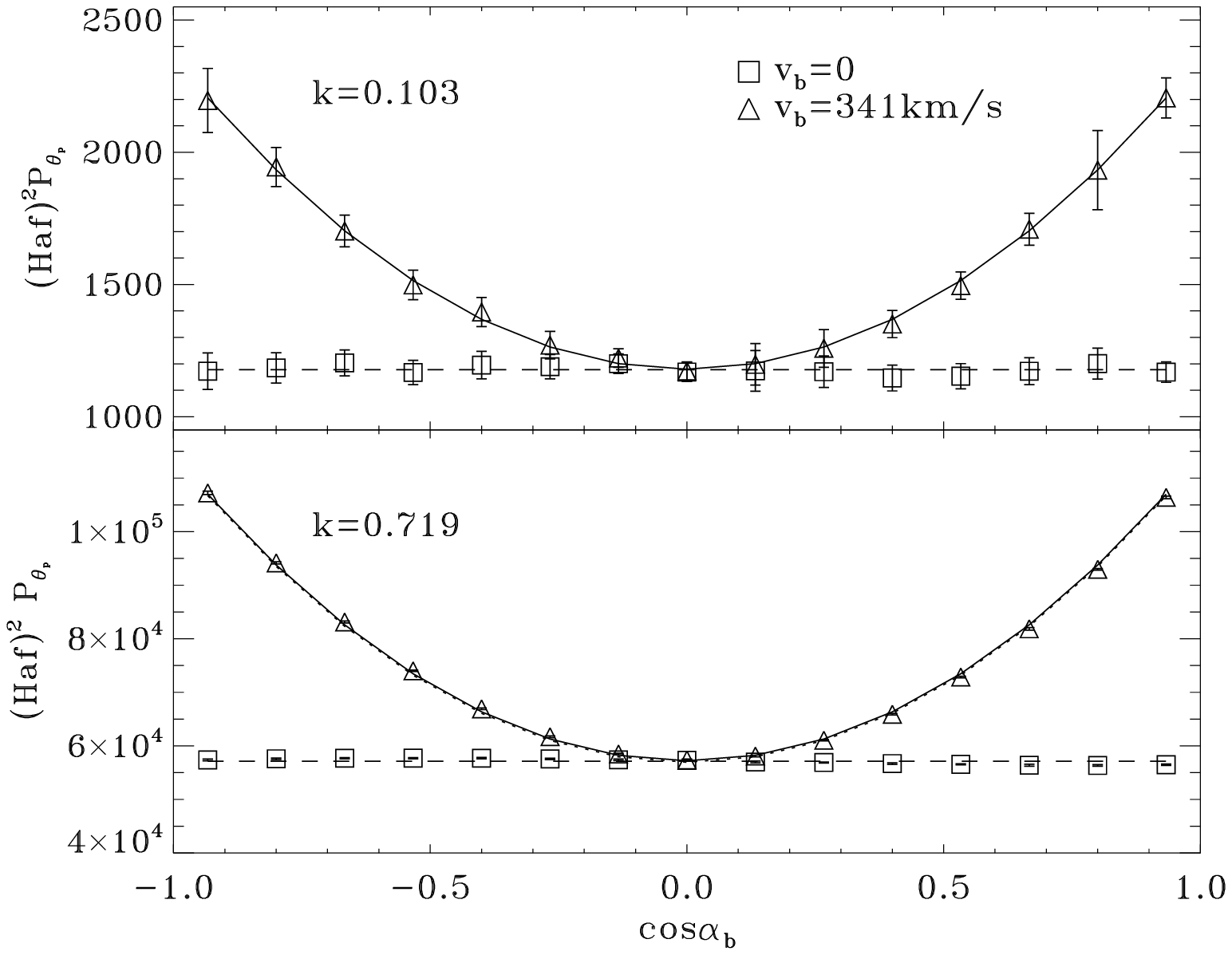}\includegraphics{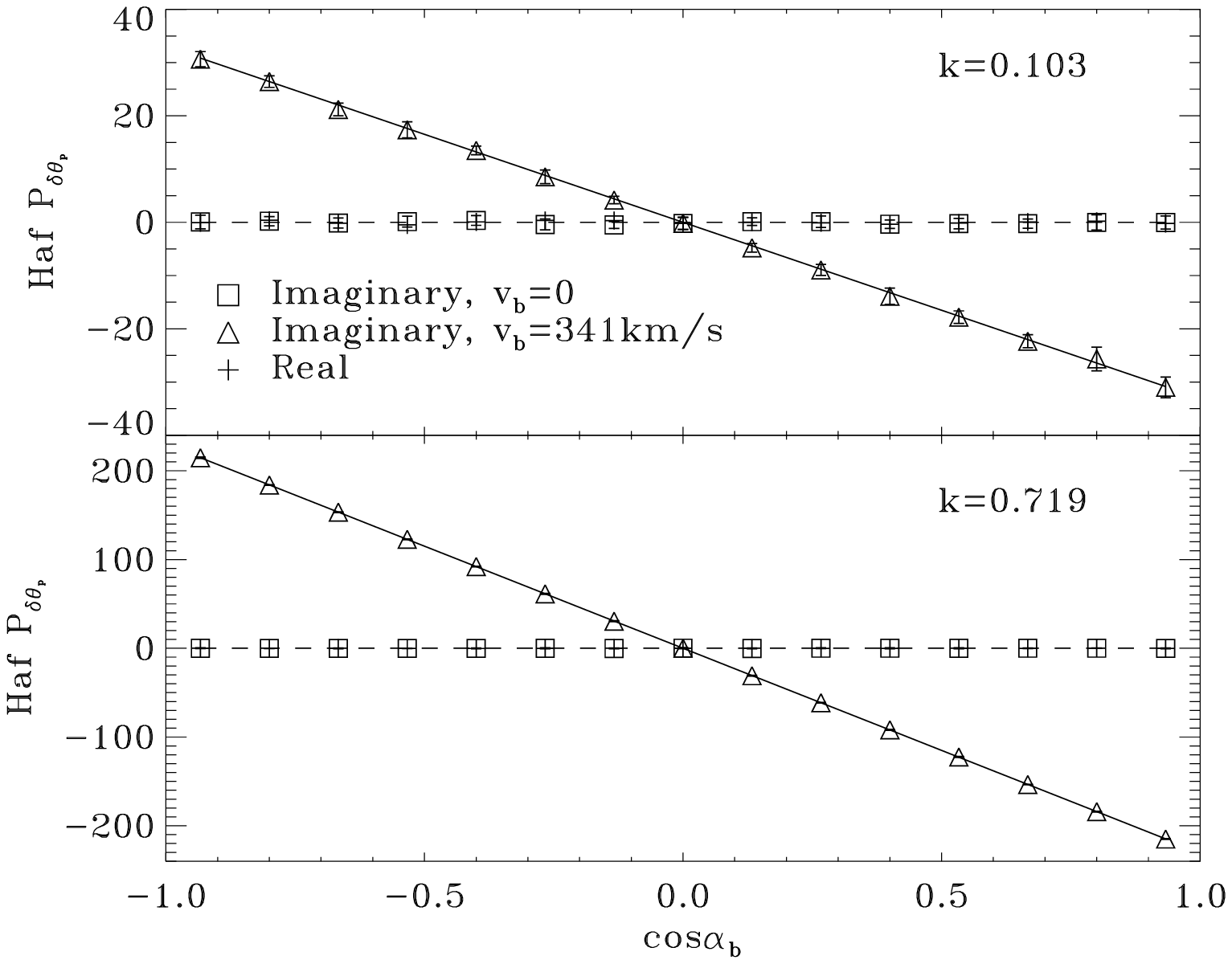}}
\caption{Raw power spectra of random samples, displayed as test of shot noise models.
Symbols are averages of ten random samples, error bars are their standard deviations, 
lines are expectation of shot noise models. $\alpha_b$ is the angle between 
bulk flow $\vect{v}_b$ and wave vector $\vect{k}$. Random samples with non-zero bulk flow
are created by adding a flow of $v_b=341$km/s along particular direction.
}
\label{fig:vbnoise}
\end{figure*}

Models of shot noise are checked with the ten random samples at $z=0$.
The randomization procedure erase any nontrivial correlations among density and velocity,  
their raw power spectra are simply signals of shot noise. Meanwhile as only particle positions are 
changed, $\sigma_v^2$ and $\avg{v^2\mu^2}$ are kept invariant, and $v_b=0$. In the experiment, 
the nearest grid point (NGP) method is used to assign objects to FFT grids, as for uniformly random
samples NGP method is exact \citep{Jing2005}. We did compare
results using the third order Bettle-Lemari\'{e} scaling function, in general 
the resulting random fluctuation is less than $1\%$, whilst systematic difference is 
around $0.4\%$ when $k$ becomes close to Nyquist frequency.

Comparison between measurements and models are presented in Figure~\ref{fig:vbnoise}, it clearly
indicates that performance of models is satisfactory, except that fluctuation of measured shot noise 
in cross-spectrum is larger than others. In order to test effects of bulk flow on shot noise, an 
artificial bulk flow of $v_b=341{\rm km/s}$ is 
added to the random samples along particular direction, and then power 
spectra are measured for comparison. By Figure~\ref{fig:vbnoise}, it is clear the shot 
noise models are indeed working very well.

%--------------------------------------------------------------------------------------------------
\subsubsection{Aliasing}

\begin{figure*}
\resizebox{\hsize}{!}{\includegraphics{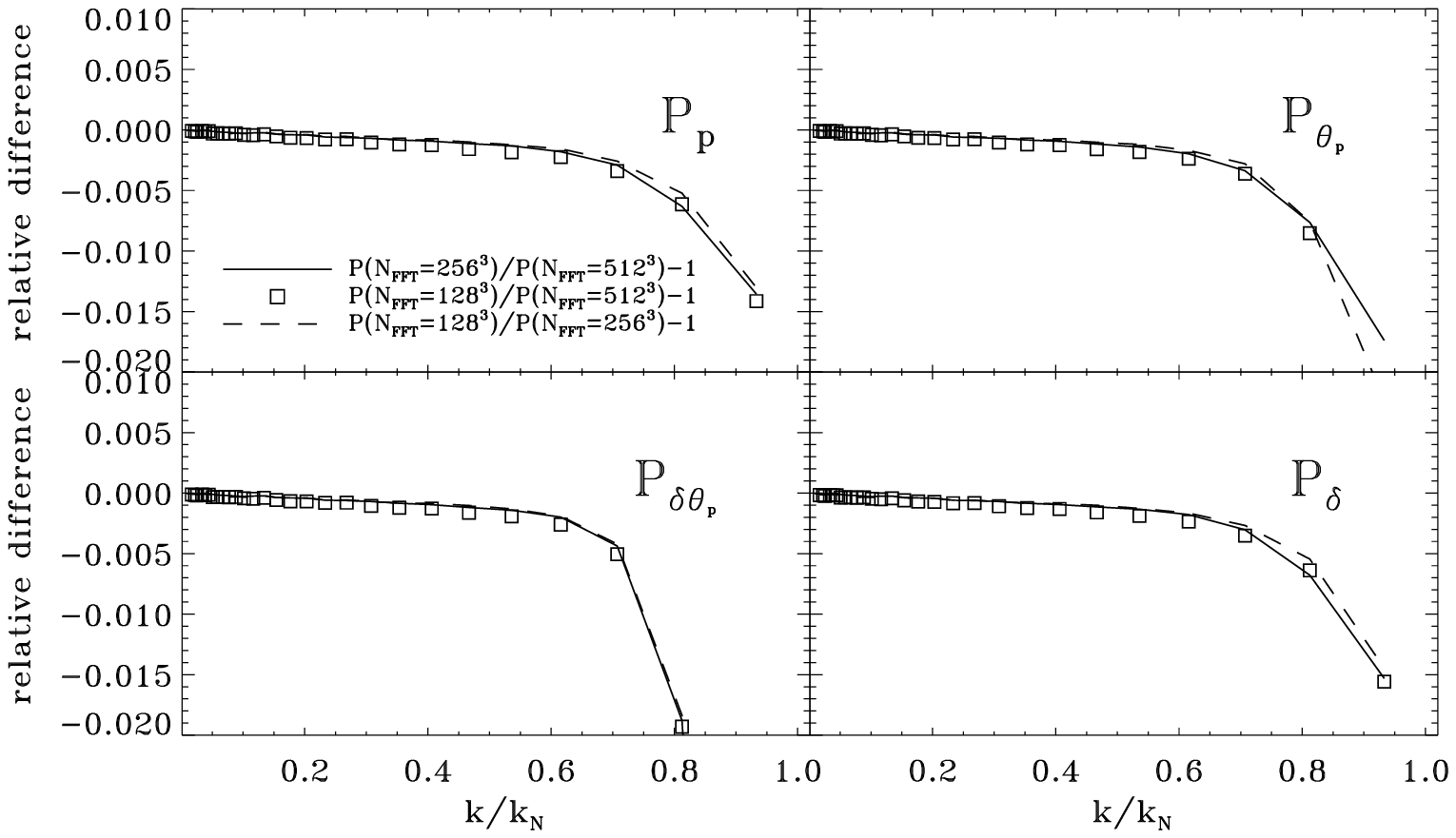}}
\caption{Aliasing effects in power spectra of dark matter at $z=0$ which object assignment 
upon FFT grids is realized with the third order Bettle-Lemari\'{e} scaling function.  
$N_{FFT}$ is the number of grids used for FFT, $k_N$ is the Nyquist frequency of the measurement
of lower FFT resolution in each pair of power spectra for comparison.}
\label{fig:simalias}
\end{figure*}

Aliasing effects in power spectra are checked with the dark matter sample at $z=0$ of N-body simulation, 
the third order Bettle-Lemari\'{e} scaling function is adopted to assign objects upon FFT grids. 
As there is no the {\it true} power spectra as template for comparison, power spectra are estimated
with different resolutions of FFT grids, then measurements of low FFT resolutions are compared with those of
higher FFT resolutions (Figure~\ref{fig:simalias}).  
It appears that the performance of the algorithm is fairly
satisfactory, the consistency indicates that for $k$ scales below the Nyquist frequency $k_N$ the aliasing 
damping to the power spectrum is tiny:
\begin{itemize}
\item the $k$ scale above which deviations are larger than $1\%$ are
about $0.87k_N$, $0.82k_N$, $0.74k_N$ and $0.85k_N$ for $P_p$, $P_{\theta_p}$, $P_{\delta\theta_p}$
and $P_\delta$ respectively;
\item at scales $k\leq 0.7k_N$ precision of $0.5\%$ can be ensured, while at scale of $0.5k_N$, the relative differences are less than $0.2\%$;
\item even when $k$ is very close $k_N$ relative differences are generally
less than $5\%$. 
\end{itemize} 
As a reference, if NGP is used for object assignment, scales where difference
is larger than $1\%$ are $\sim0.12k_N$ for these power spectra, the aliasing damping is
much severe.

%%=======================================================================
\section{Impact of subtracting bulk flow}
In the previous section the influence of non-zero bulk flow on the shot noises of the momentum power spectra
has been analyzed. However effects of non-zero bulk flow could be more than the simple modulation to shot noises. \citet{ParkPark2006} noticed the problem, and \citet{Howlett2019} carried out extensively
numerical exploration with mock catalogues, they conjectured 
that removing the bulk flow from measured peculiar velocities brings little changes to power spectra,
but reminded that in practical works it should be tested case by case. 
In this section we will mainly focus on the changes after subtracting bulk flow from peculiar velocities 
to the estimated momentum power spectra. 

%%-----------------------------------------------------------------------------------------------------------------------------
\subsection{Momentum}
Since aliasing can be well corrected in our algorithms, in the following derivation we will not include
aliasing effects any longer. In the case of the bulk flow $\vect{v}_b=\sum_j \vect{v}_j/N \neq 0$, it is 
always possible to define a new velocity by removing the bulk flow $\vect{v}'=\vect{v}-\vect{v}_b$ to 
generate a new
momentum field $\vect{p}'=\vect{p}-(1+\delta)\vect{v}_b$ with zero bulk flow.
The raw power spectrum of the new momentum field is 
\begin{equation}
\widehat{P}_{p'}(\vect{k})=  \frac{1}{N^2} \sum_{j \neq \ell} 
\avg{ (n_j \vect{v}'_j)\cdot (n_\ell \vect{v}'_\ell)}  e^{i \vect{k} \cdot (\vect{r}_j - \vect{r}_\ell)} 
+ \frac{\hat{\sigma}_{v'}^2}{N} \ ,
\label{eq:rawppb}
\end{equation}
which shot noise is related to Eq.~\ref{eq:rawppnoise} through
\begin{equation}
\hat{\sigma}^2_v=\sum_{j, n_j=1}{v'}^2_j/N+v_b^2=\hat{\sigma}^2_{v'}+v_b^2\ .
\end{equation}
Correlation functions in Eqs.~\ref{eq:calsn} and~\ref{eq:rawppb} are linked by
\begin{equation}
\avg{n_j\vect{v}_j\cdot n_\ell \vect{v}_\ell}_{j\neq\ell}=
\avg{n_j\vect{v}'_j\cdot  n_\ell \vect{v}'_\ell}_{j\neq \ell} 
+v_b \avg{n_j n_\ell  (v'_j\eta'_j+v'_\ell \eta'_\ell)}_{j\neq\ell} 
+ v_b^2\avg{n_jn_\ell}_{j\neq \ell} \ ,
\end{equation}
in which $v'_j=\abs{\vect{v}'_j}$, $v_b=\abs{\vect{v}_b}$ 
and $\eta'_j= \vect{v}'_j\cdot \vect{v}_b/ (v'_jv_b)$.
Note that there is the correspondence
\begin{equation}
\begin{aligned}
\avg{(n_j \vect{v}_j)\cdot (n_\ell \vect{v}_\ell)}_{j\neq \ell} & \leftrightarrow
P_p(\vect{k}) \\
\avg{(n_j \vect{v}'_j)\cdot (n_\ell \vect{v}'_\ell)}_{j\neq \ell} & \leftrightarrow
P_{p'}(\vect{k})\\
\avg{n_j n_\ell  (v'_j\eta'_j+v'_\ell \eta'_\ell)}_{j\neq\ell} & \leftrightarrow
P_{\delta p'_b}(\vect{k})+P_{\delta p'_b}(\vect{-k}) \\
\avg{n_jn_\ell}_{j\neq \ell} & \leftrightarrow P(\vect{k}) \ ,
\end{aligned}
\end{equation}
where $P(\vect{k})=\avg{\delta(\vect{k})\delta^*(\vect{k})}$, 
$P_{\delta p'_b}(\vect{k})=\avg{\delta(\vect{k}) p'^*_b(\vect{k})}$,
$p'_b({\vect{k}})$ is the Fourier transform of 
$p'_b(\vect{r})=\rho \vect{v}' \cdot \vect{v}_b/v_b=\rho v'\eta'$, finally there is the
the relation
\begin{equation}
P_p(\vect{k})=P_{p'}(\vect{k})+
v_b\left[ P_{\delta p'_b}(\vect{k})+P^*_{\delta p'_b}(\vect{k}) \right] +v^2_b P(\vect{k})\ .
\label{eq:bfpp}
\end{equation}
In practical application, $P_p$ and $P_{p'}$ can be estimated via
Eqs.~\ref{eq:rawPp} , $P(\vect{k})$ can be
measured through $\widehat{P}(\vect{k})=P(\vect{k})+1/N$, while
$\widehat{P}_{\delta p'_b}=P_{\delta p'_b}$.

%--------------------------------------------------------------------------------------------
\subsection{Momentum divergence}
The quantity implemented in algorithm to estimate statistics of momentum divergence is constructed by 
\begin{equation}
\hat{\theta}_p({\vect{k}})=\frac{1}{N}\sum_j \frac{i \vect{k}}{Haf} \cdot (n_j \vect{v}_j) e^{i\vect{k}\cdot \vect{r}_j}\ .
\end{equation}
If $v_b\neq 0$, with $\mu_b=\vect{k}\cdot \vect{v}_b/(kv_b)$, there are
\begin{equation}
\begin{aligned}
&\avg{[i\vect{k}\cdot (n_j \vect{v}_j)] [-i\vect{k}\cdot (n_\ell \vect{v}_\ell)]}_{j\neq \ell} = 
 \avg{[i\vect{k}\cdot (n_j \vect{v}'_j)] [-i \vect{k}\cdot (n_\ell \vect{v}'_\ell)]}_{j\neq \ell} \\
&\ \ \ \ \ \ \ \ \ \ \ \ \ \ \ \ \ \ + i k v_b \mu_b 
\avg{
n_j  [-i \vect{k}\cdot (n_\ell \vect{v}'_\ell)] - [i \vect{k}\cdot (n_j \vect{v}'_j)] n_\ell}_{j\neq \ell}
   +  
  k^2 v_b^2 \mu^2_b \avg{n_j  n_\ell}_{j\neq \ell} \\
&\avg{n_j [-i\vect{k}\cdot (n_\ell \vect{v}_\ell)]}_{j\neq \ell }= 
\avg{n_j [-i\vect{k}\cdot (n_\ell \vect{v}'_\ell)]}_{j\neq \ell } -ikv_b\mu_b\avg{n_jn_\ell}_{j\neq l}\ .
\end{aligned}
\end{equation}
Subsequently if let the divergence of momentum after subtracting $\vect{v}_b$ be $\theta'_p$,
we obtain the following equations,
\begin{equation}
\begin{aligned}
& P_{\theta_p}(\vect{k})  =P_{\theta'_p}(\vect{k})+i k \mu_b \frac{v_b}{Haf}[P_{\delta\theta'_p}(\vect{k}) 
- P^*_{\delta\theta'_p}(\vect{k}) ]   + k^2 \mu_b^2 \left( \frac{v_b}{Haf}\right)^2 P(\vect{k}) \\
&P_{\delta\theta_p}(\vect{k}) =  P_{\delta\theta'_p}(\vect{k})- i k \mu_b \frac{v_b}{Haf} P(\vect{k})\ .
\end{aligned}
\label{eq:bfpdpx}
\end{equation}
$P_{\theta_p}$, $P_{\theta'_p}$, $P_{\delta\theta_p}$ and $P_{\delta\theta'_p}$ can be measured
by Eq.~\ref{eq:rawPtheta} and ~\ref{eq:rawPx} respectively.

%%----------------------------------------------------------------------------------------------------------------------------
\subsection{Numerical experiments}

\begin{figure*}
\resizebox{\hsize}{!}{\includegraphics{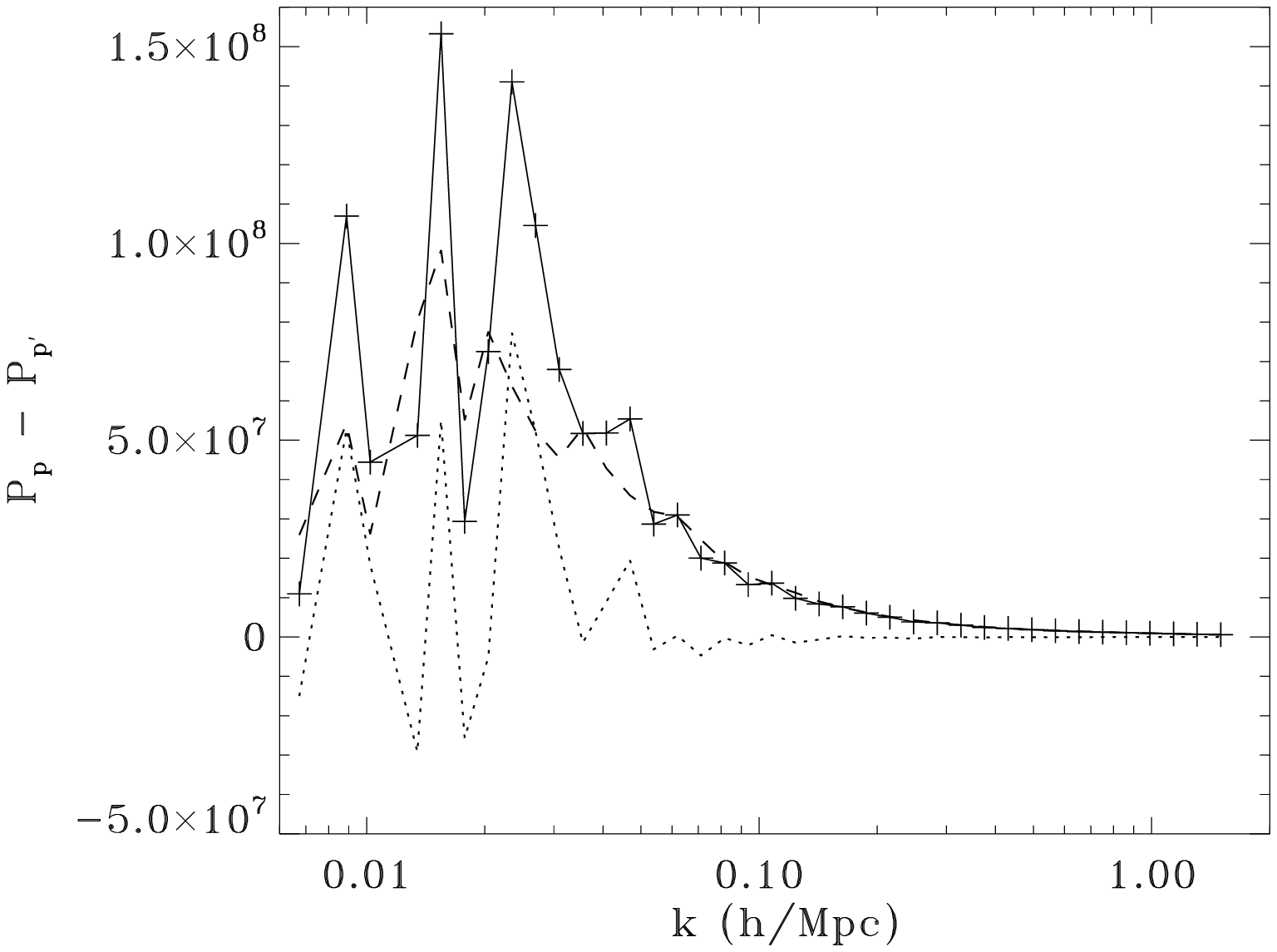}\includegraphics{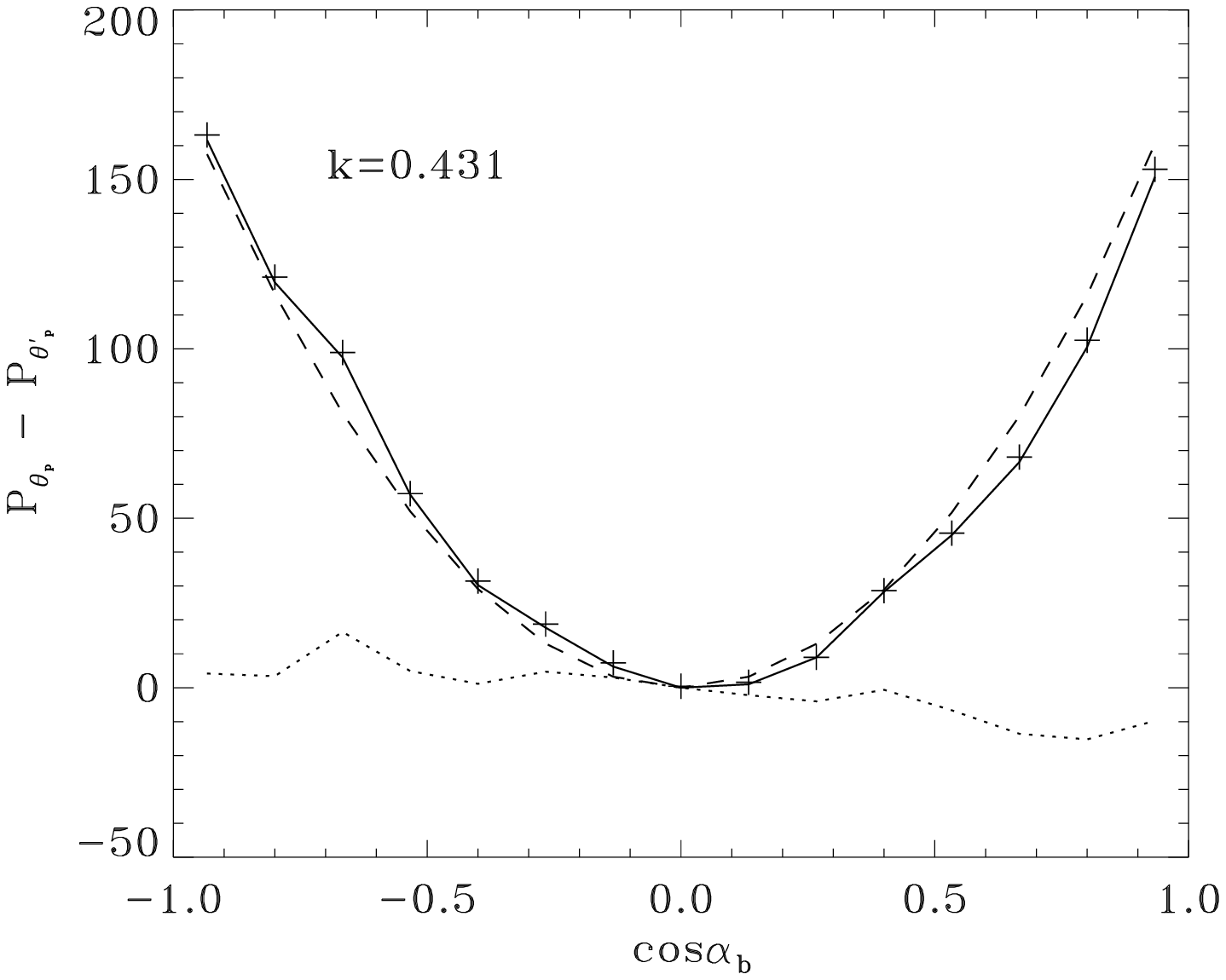}}
\resizebox{\hsize}{!}{\includegraphics{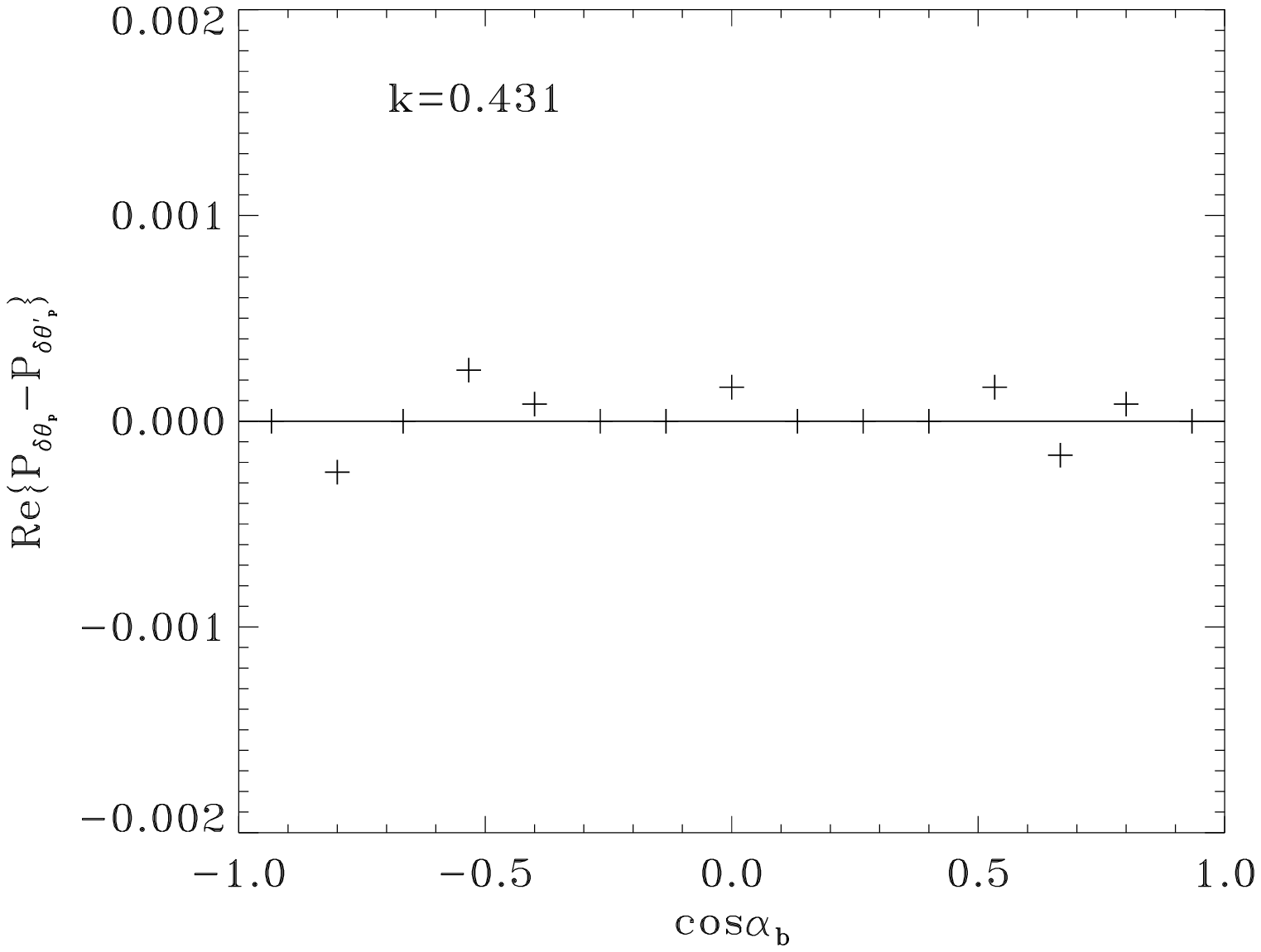}\includegraphics{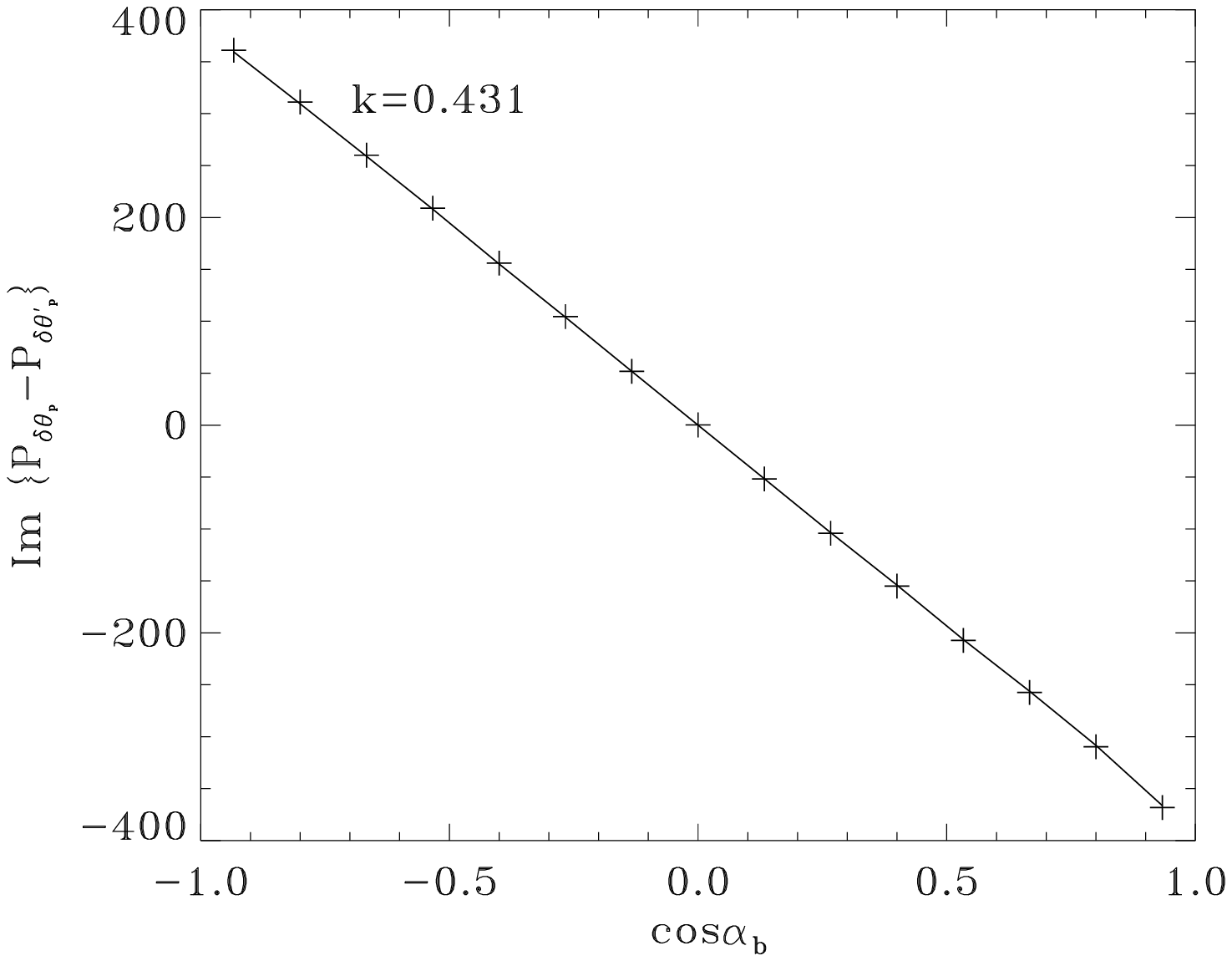}}
\caption{Differences in measured momentum power spectra after removing bulk flow.
An artificial bulk flow of speed $v_b=51.5$km/s is added to the $z=0$ realization of the
simulation to create a sample with bulk flow, $\alpha_b=\cos^{-1}\mu_b$ is the angle 
between $\vect{k}$ and $\vect{v}_b$.
Top left: cross symbols are $P_p-P_{p'}$, the solid line is 
$v_b(P_{\delta p'_b} + P^*_{\delta p'_b})+v_b^2P$, dotted line is $v_b (P_{\delta p'_b} + P^*_{\delta p'_b})$, 
and dashed line is $v_b^2P$ (Eq.~\ref{eq:bfpp}).
Top right: crosses are $P_{\theta_p}-P_{\theta'_p}$, solid line is 
$ik \mu_b (P_{\delta\theta'_p}-P^*_{\delta}) v_b/(Haf)+k^2\mu_b^2 (v_b/Haf)^2 P$ in which
the first term is drawn in dotted line and the second term is the dashed line (Eq.~\ref{eq:bfpdpx}).
Bottom left: crosses are the real part of $P_{\delta\theta_p}-P_{\delta\theta'_p}$, solid line
is the expectation of zero. Bottom right: the imaginary part of $P_{\delta\theta_p}-P_{\delta\theta'_p}$,
solid line is $-k\mu_b (v_b/Haf) P$ (Eq.~\ref{eq:bfpdpx}). }
\label{fig:bfPpdiff}
\end{figure*}

It is well known that bulk flow of a sample follows Maxwellian distribution 
\begin{equation}
\mathcal{P}(v_b)\rmd v_b =\sqrt{\frac{2}{\pi}} \left( \frac{3}{\sigma_{v_b}^2} \right)^{3/2} v_b^2 \exp \left( -  \frac{3v_b^2}{2\sigma_{v_b}^2} \right) \rmd v_b
\label{eq:vbpdf}
\end{equation}
which is
solely controlled by the variance 
$\sigma_{v_b}^2=\int P_{\vect{v}}\widetilde{W}_S^2 d^3k/(2\pi)^3$, 
$\widetilde{W}_S$ is the window function defining the sample space 
in Fourier space \citep[e.g.][]{BahcallEtal1994, LiEtal2012}. Usually the
sample space is sufficiently large to approximate the mass-weighted velocity
power spectrum $P_{\vect{v}}$ with the linear power spectrum of 
density fluctuation $P_L$ by $(Haf/k)^2P_L$. The most likely speed of
bulk flow is $\sqrt{2/3} \sigma_{v_b}$, the mean 
$\avg{v_b}= \sqrt{8/(3\pi)}\sigma_{v_b}$ and the mean square speed 
$\avg{v_b^2}=\sigma^2_{v_b}$. 
The characteristic speed of bulk flow corresponding to the volume 
of our simulation is $\sigma_{v_b}=51.5$km/s, 
so an artificial bulk flow $v_b=\sigma_{v_b}$ in an arbitrary selected direction
is added to the dark matter sample at $z=0$ of the simulation to form a sample with non-zero
bulk flow, then power spectra are estimated to check the resulting influence. 
Summary of our experiments is shown in Figure~\ref{fig:bfPpdiff}.
Eqs.~\ref{eq:bfpp} and Eq.~\ref{eq:bfpdpx} are confirmed with excellent 
accuracy (better than $0.03\%$), 
even the statistical fluctuations due to limited number of modes at large scales are recovered perfectly.

\begin{figure*}
\resizebox{\hsize}{!}{\includegraphics{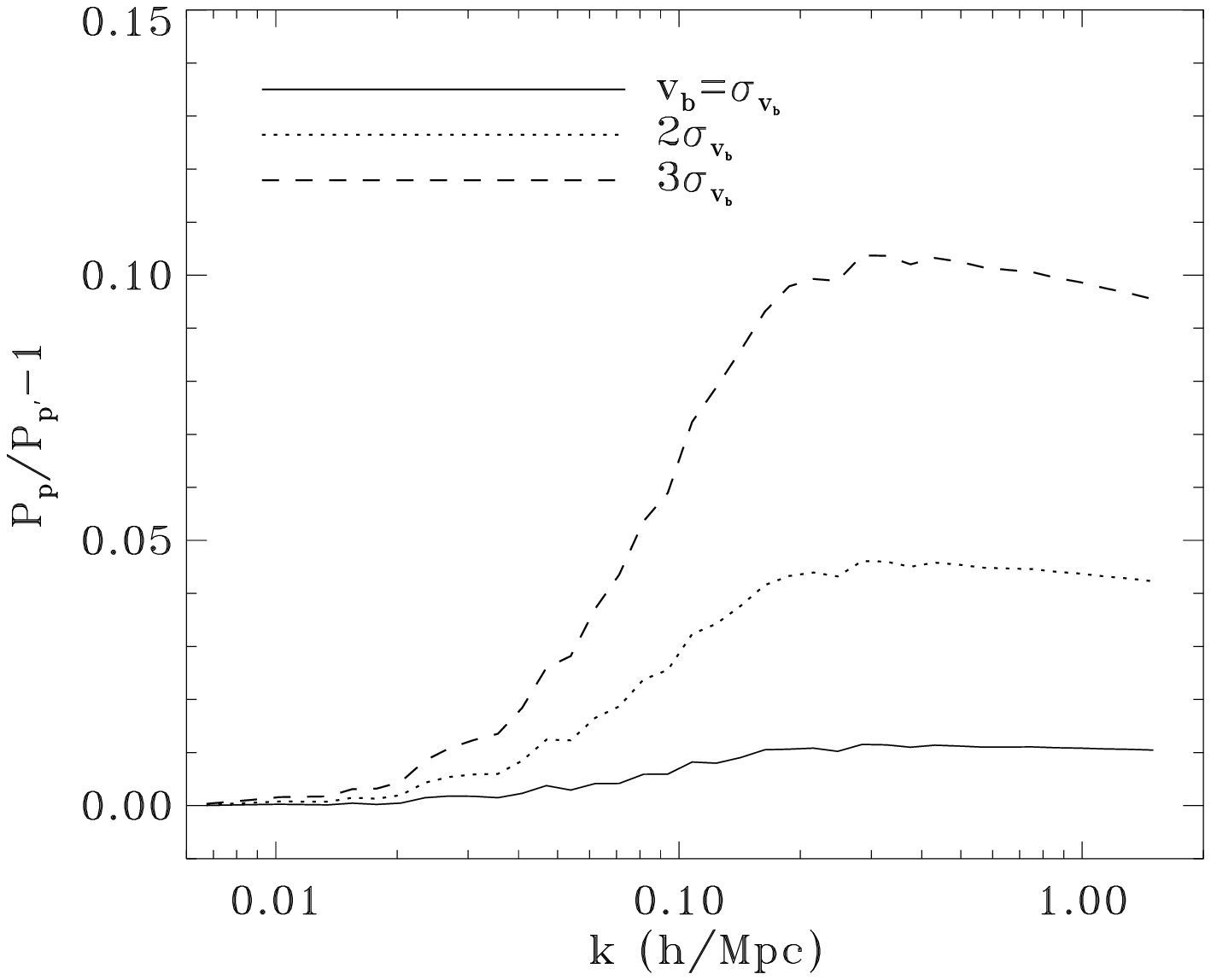}\includegraphics{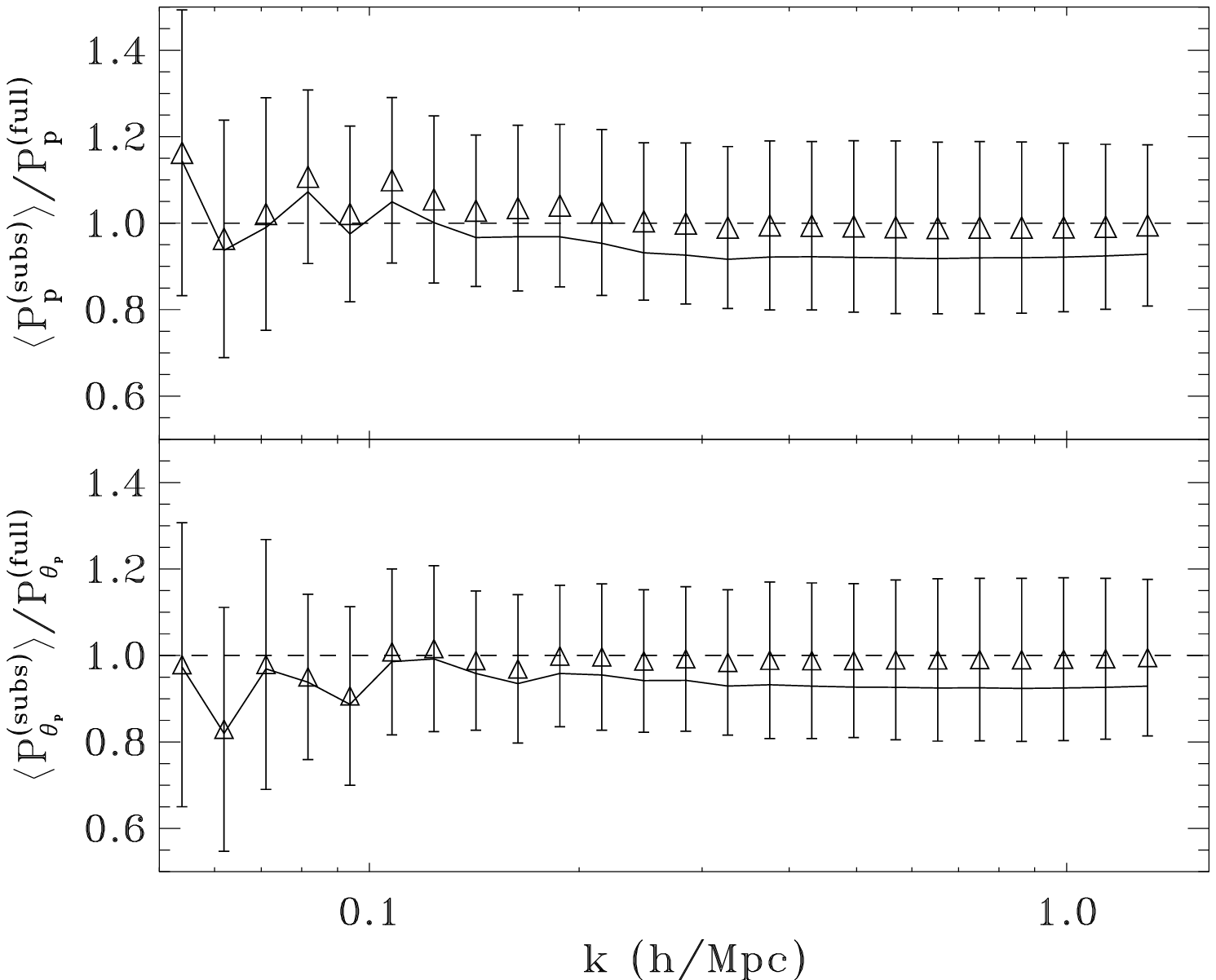}}
\caption{Left panel: differences in $P_p$ after removing bulk flows of different amplitudes,
$\sigma_{v_b}=51.5$km/s. Right panel: differences between averages of power spectra of
the 64 subsamples and power spectra of the full sample; triangles are averages of power spectra of 
subsample without removing their bulk flows, error bars are their standard deviations; 
solid lines are the averages of the subsamples' power
spectra with their bulk flows subtracted; dashed horizontal lines are the reference lines of $1$.}
\label{fig:bfppdiff_vb}
\end{figure*}

It is apparent the modulation 
depends on the amplitude of $v_b$, major contribution comes from $v_b^2P$, 
at large scales $P_p\sim (Haf)^2 k^{-2}P$, the relative difference
$P_p/P_{p'}-1$ is roughly $k^2 v^2_b/(Haf)^2 \sim 4 (v_b/100)^2 k^2$ 
at $z\sim 0$.
If sample volume is large, the chance to have a large bulk flow is relatively small, it is
expected that $P_p$ at large scales will not change significantly by subtracting the bulk flow, but at small 
scales one might have to consider the difference, as shown in the left panel of Figure~\ref{fig:bfppdiff_vb}. 
An important issue one has to bear in mind, what is presented in Figure~\ref{fig:bfppdiff_vb} is of dark matter. 
If at large scales $k< \sim 0.1$h/Mpc, peculiar velocities of biased objects such as galaxies are only 
slightly biased with respect to the dark matter, i.e. $b_v\approx 1$ \citep{ChenEtal2018},  
the correction to the momentum 
power spectra of galaxies after removing bulk flow will be actually boosted by the square of the galaxy
density bias parameter.

A serious question is whether $\vect{p}$ or $\vect{p}'$ should be used to estimate the power spectra. 
Non-zero monople of peculiar velocities can also emerge by
systematics in peculiar velocity estimation methods, i.e. the velocity zero point offsets which 
might have distinct distribution function from the intrinsic flow.
If the measured bulk flow is caused by the peculiar velocity zero point offset alone, 
no doubt that one needs to deduct the measured bulk flow directly. 
If peculiar velocities are given exactly, such as in samples constructed from simulation data,
the bulk flow is purely intrinsic, it is then another story. In order to clarify the point,
power spectra of our 64 subsamples at $z=0$ are estimated with and without their particular bulk 
flow subtracted respectively. 
Then averages of these power spectra are compared with the measurements of the sample of full 
size to check possible systematical biases. As expected by Eqs.~\ref{eq:bfpp} and ~\ref{eq:bfpdpx}
after replacing $v_b^2$ with $\sigma_{v_b}^2$,
we can see from the right panel of Figure~\ref{fig:bfppdiff_vb}, that subtracting bulk flows from the subsamples
gives rise to systematically biased estimation of $P_p$ and $P_{\theta_p}$ at small
scales, although such biases seem not so significant against
the fairly large dispersions among the measured momentum power spectra
of subsamples. Nevertheless it appears that there is no need to subtract the bulk flow in this case.
Measured bulk flow of real samples contains mingled contributions from both of the intrinsic flow and 
the velocity zero point offsets, one might have to inspect the strengths of the two sources carefully case
by case.

%%=======================================================================
\section{Momentum power spectra of dark matter in the $\Lambda$CDM simulation}
\subsection{At large scales}

\begin{figure*}
\center{\resizebox{0.67\hsize}{!}{\includegraphics{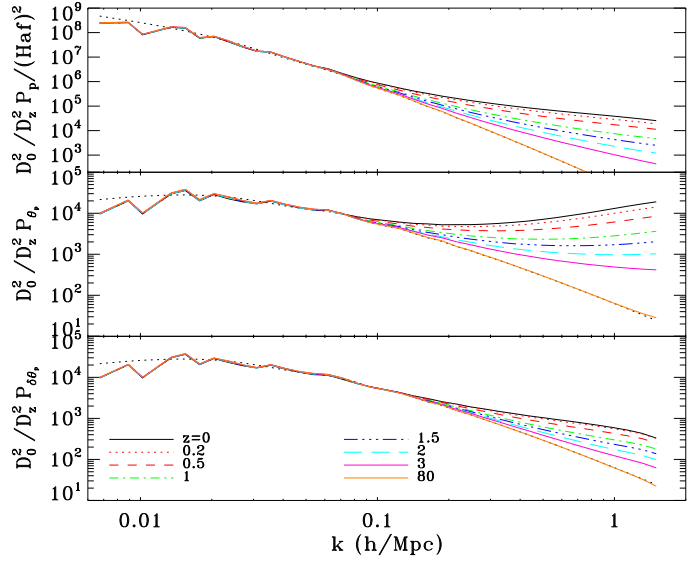}}}
\caption{Power spectra of dark matter in our $\Lambda$CDM simulation at selected redshifts
from $z=0$ to the simulation's starting epoch $z=80$, black dotted lines which are almost coincident
with the measurements at $z=80$ are of the linear theory. 
}
\label{fig:simsall}
\end{figure*}

As an application, momentum power spectra of dark matter in the $\Lambda$CDM simulation at many epochs 
from the initial time of $z=80$ to $z=0$ are estimated with our algorithms. Measured power spectra of the
full simulation are shown in Figure~\ref{fig:simsall}, it looks that linear theory matches
simulations at large scales well, but the scale ranges allowed by the simulation for 
accuracy examination are very narrow, the box size of our simulation is $1$Gpc/h which in 
Fourier space corresponds to $k\approx 0.006$, the strong fluctuation at large scales 
in power spectra caused by limited number of Fourier modes becomes an obstacle to observe
the actual performance of theories.

\begin{figure*}
\resizebox{\hsize}{!}{\includegraphics{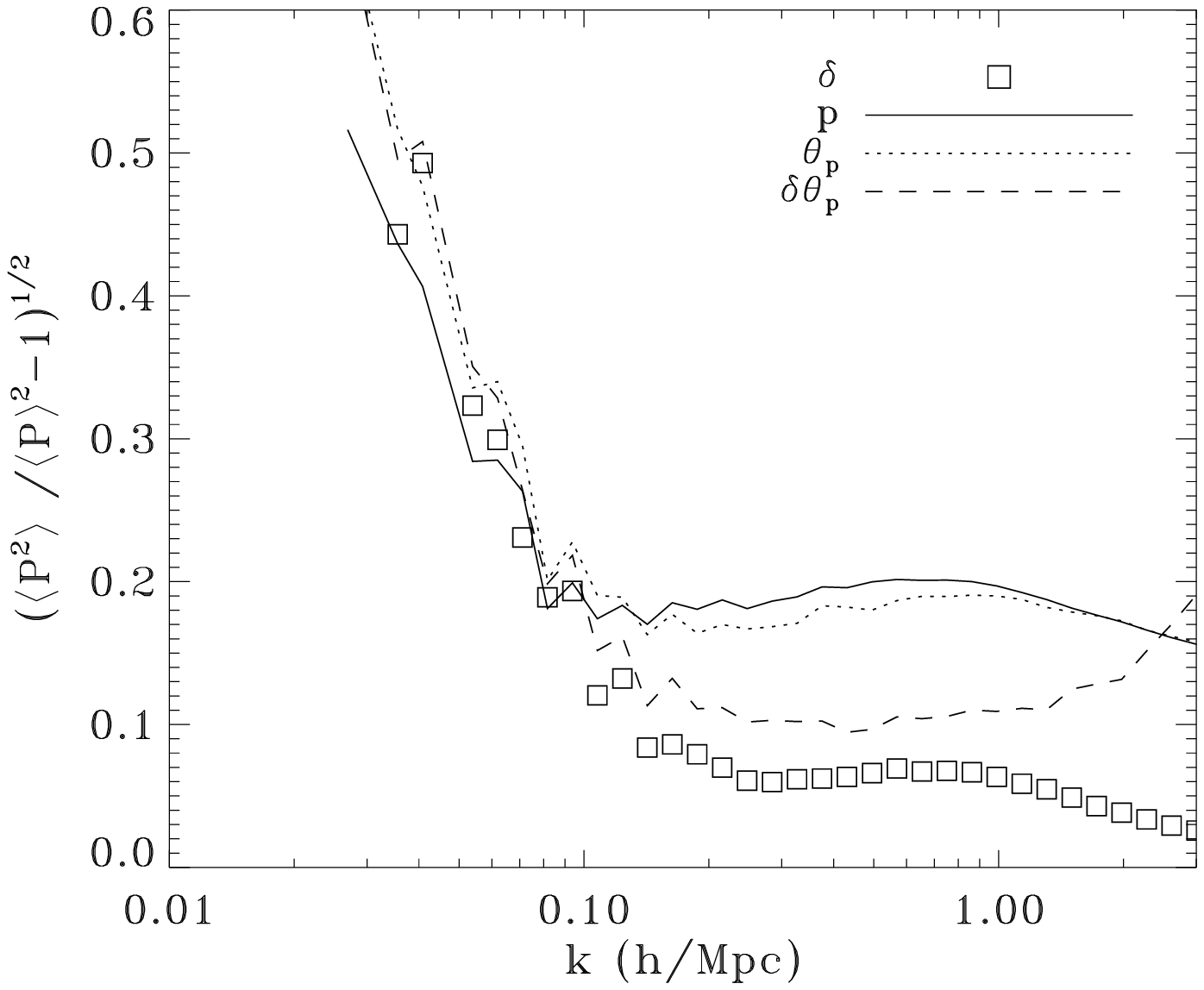}\includegraphics{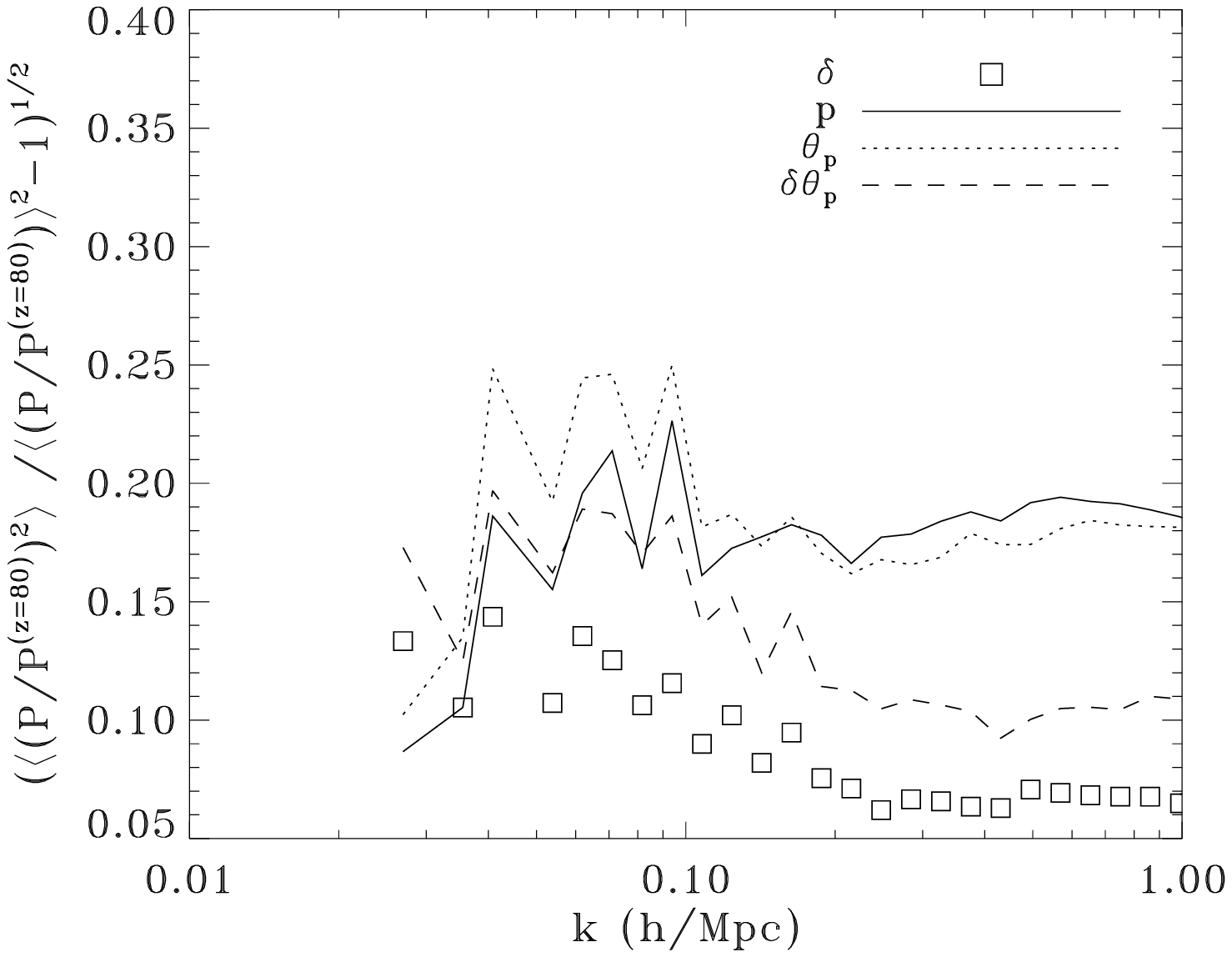}}
\caption{Left: relative uncertainties in power spectra of dark matter at $z=0$, being
the normalized standard deviations among measurements of 64 subsamples, which are
also the relative uncertainties of the ratios of power spectra to the linear theory. Right:
the relative uncertainties of $P/P^{(z=80)}$, $P$ refers to power spectrum of the specific kind at $z=0$,
$P^{(z=80)}$ is the linear template as the power spectrum estimated from corresponding subsample
extracted from the initial field (at $z=80$) and linearly evolved to $z=0$.}
\label{fig:var}
\end{figure*}

Considering that we have only one simulation at hand, we estimated error bars as the standard deviation of the
measurements of the 64 subsamples of the $z=0$ output used in last section, the shortcoming of this 
method is that since the box size of subsamples is only one quarter of the original full sample, error bars below
$k< \sim 0.025$ are missing in our application thereof. We can see that uncertainties at large scales 
$k< \sim 0.1h/$Mpc are quite large (left panel of Figure~\ref{fig:var}), which is known to be
roughly inversely proportional to the square root of numbers of Fourier modes. Variances of momentum 
power spectra are persistently several times stronger than that of density power spectrum at $k>\sim 0.1h$, 
being around $20\%$ of $P_p$ and $P_{\theta_p}$, $\sim 10\%$ of $P_{\delta\theta_p}$.

To assess precision of theoretical models, the object quantities estimated from simulation should 
contains stochastic fluctuation as less as possible, overlaying error bars on the estimated power spectra 
only indicate range of uncertainties, a method able to suppress sample variance would be very helpful. 
We realize that at very large scales, 
coupling among Fourier modes is in fact weak, Fourier modes can be deemed evolving
linearly, such that data sets at later epochs actually maintain 
approximately the same large scale stochastic fluctuations as the random setup in 
the initial condition. We take the measured power spectra of the initial field granted as the 
linear templates, which differ from linear theoretical models by less than $2\%$ if 
ignoring the cosmic variance. Thereafter using these linear templates to normalize 
measured power spectra at later times shall be able to alleviate cosmic variances. 

\begin{figure*}
\resizebox{\hsize}{!}{\includegraphics{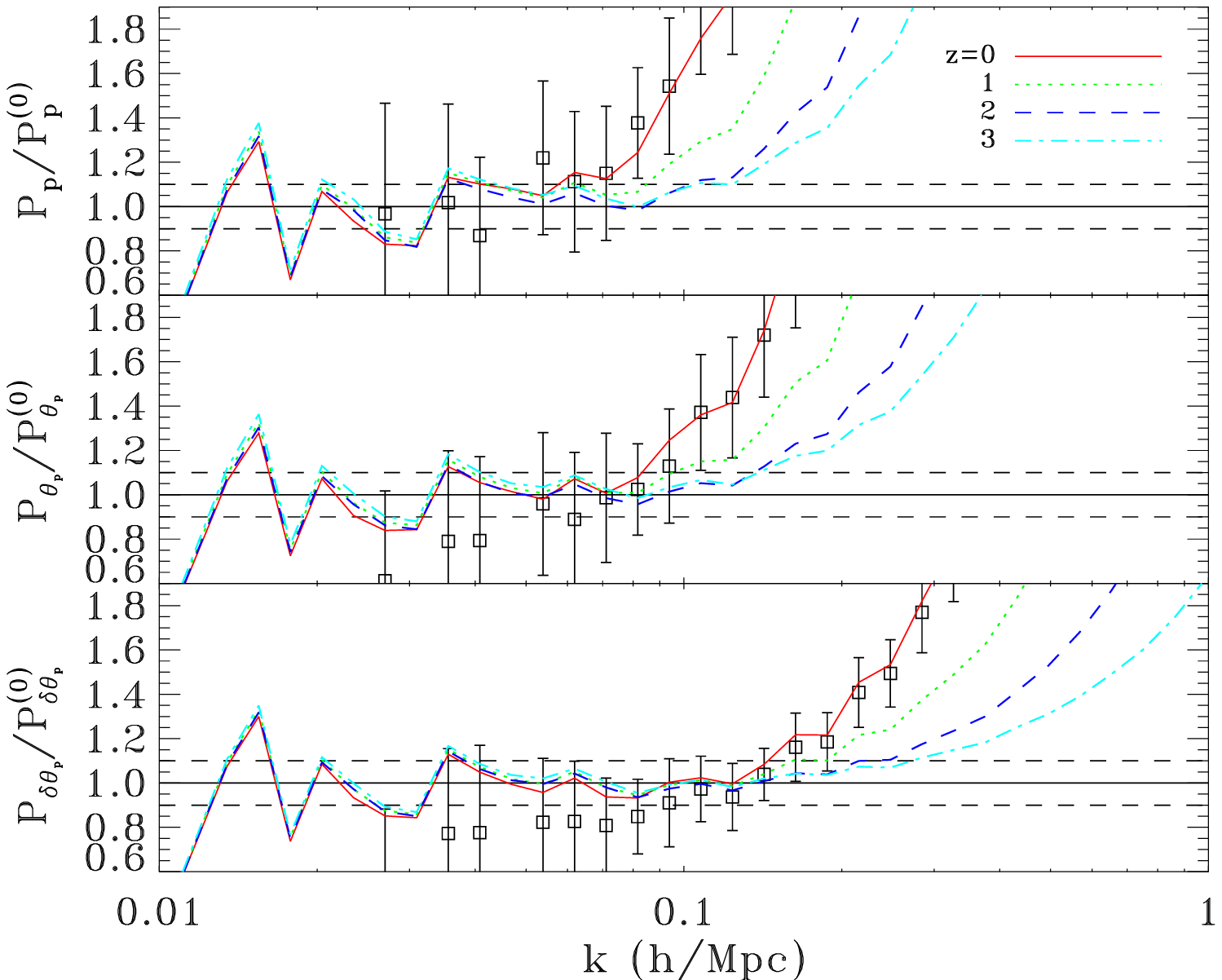}\includegraphics{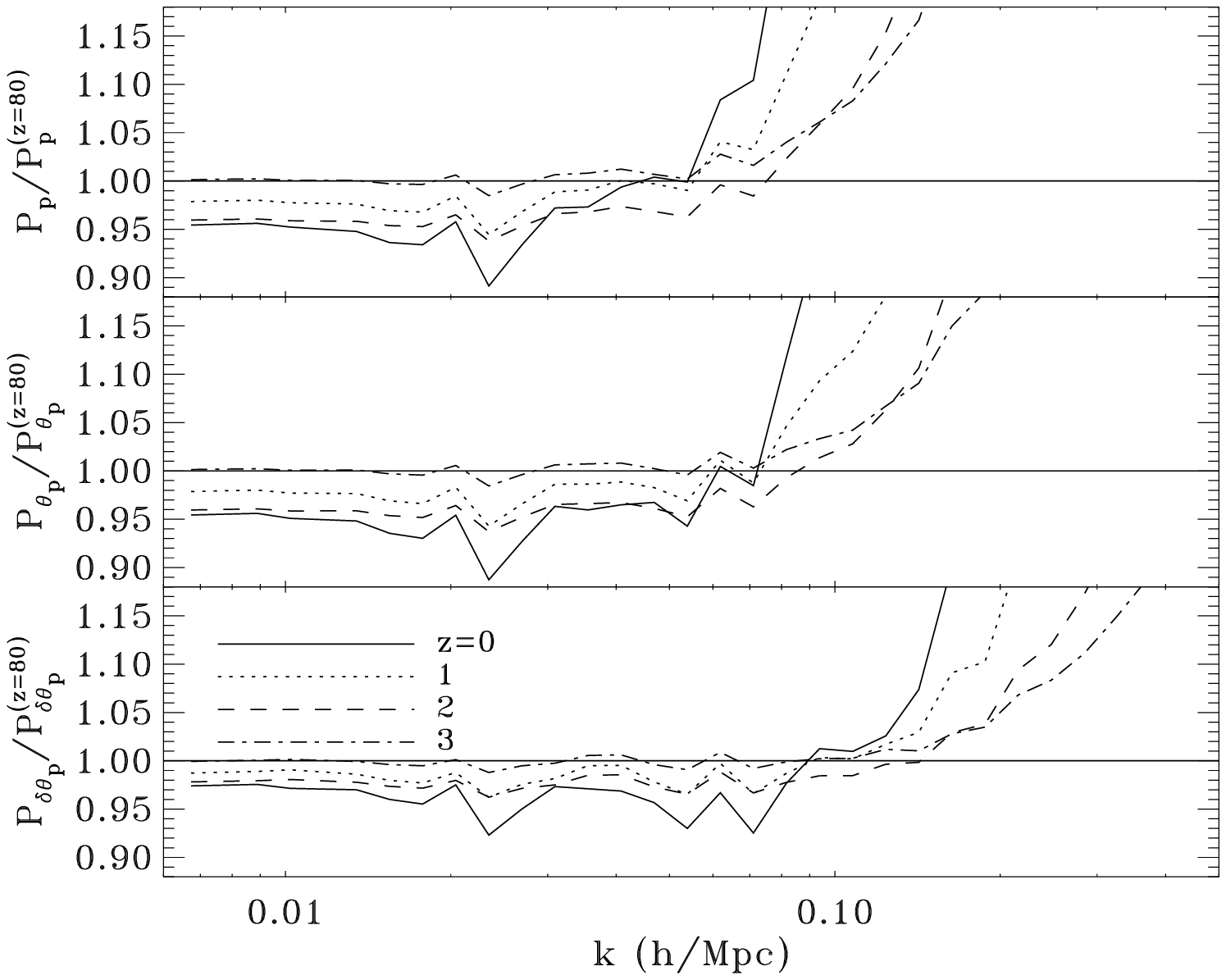}}
\caption{Left panel: ratios of power spectra measured from
simulation to the linear theory, square symbols are the averages of the 64 subsamples while error bars are
corresponding standard deviations, solid horizontal lines are the unity ratio and the dashed horizontal lines
delimit  the $10\%$ deviation; linear theory predicts that
$P_p^{(0)}=D_z^2(Haf/k)^2P_0$, $P_{\theta_p}^{(0)}=P_{\delta\theta_p}^{(0)}=D_z^2P_0$,
$P_0$ is the theoretical linear density power spectrum scaled to $z=0$. Right:
measured power spectra of the full sample, after normalization by linear templates.
}
\label{fig:sim2lin}
\end{figure*}

To check the conjecture, power spectra of the 64 subsamples of the initial condition at $z=80$ are 
then measured and linearly evolved to redshift $z=0$, forming the class of linear templates denoted 
as $P^{(z=80)}$. 
Uncertainties are then estimated for power spectra normalized by these linear templates. 
The technique is 
indeed very effective, dramatically reduces the cosmic variances at 
large scales $k< 0.1h$/Mpc (Figure~\ref{fig:var}), relative uncertainties in momentum power spectra
drop to $\sim 20\%$ and become much stable. 
Comparison of linear theories with the measurements of 
simulation is displayed in Figure~\ref{fig:sim2lin}, the advantage of using the measured initial power spectra
as linear prediction is obvious, results are much smooth and convergent.

%%-------------------------------------------------------------------------------------------
\subsection{Beyond linear regime}
\begin{figure*}
\center{\resizebox{0.67\hsize}{!}{\includegraphics{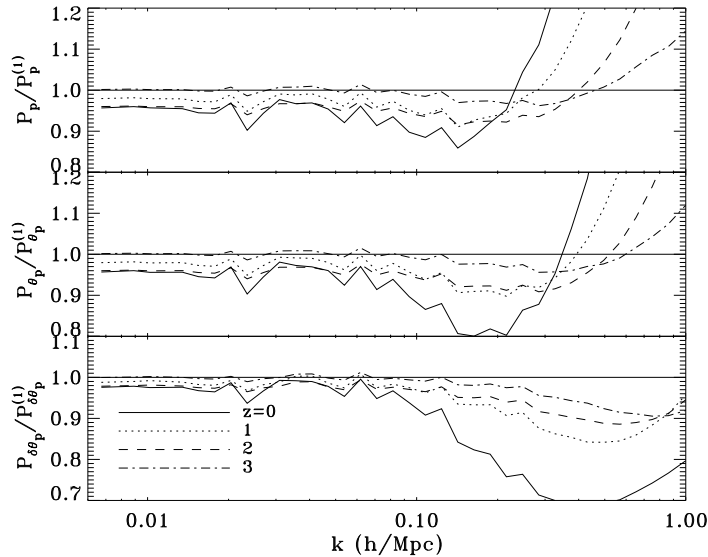}}}
\caption{Comparison of simulation with SPT at one loop level. To reduce sample variance, measured
power spectra are normalized by $P^{(z=80)}$, theoretical power spectra 
$P^{(1)}$ are normalized by $P^{(0)}$ as well. $P_p^{(1)}$ is given by Eq.~\ref{eq:PpdD4}, 
$P_{\theta_p}^{(1)}$ and $P_{\delta\theta_p}^{(1)}$ are calculated with Eqs.~\ref{eq:pdivpD4}}.
\label{fig:sim2spt}
\end{figure*}

It is not an easy task to predict nonlinear $P_p$, all nonlinear
polyspectra on the right hand side of Eq.~\ref{eq:Ppd} are needed, among which
however only the nonlinear matter power spectrum over large scale range can be 
provided with good precision by either empirical 
fitting formulae \citep{SmithEtal2003, TakahashiEtal2012} or 
halo model \citep[e.g.][]{MaFry2000a,ScoccimarroEtal2001}.
At large scales where nonlinearity is weak one could resort to
perturbative approach, such as the standard Eulerian perturbation theory (SPT, Appendix~\ref{sec:appSPT}).
The one loop approximation of SPT on momentum power spectra  (details in Appendix~\ref{sec:1loop})
is compared with simulation results in Figure~\ref{fig:sim2spt}. The one loop SPT brings
minor improvement over linear theory for the case of $z=0$, but could be applied to slightly
deeper scales at high redshifts $z>1$ if precision requirement is as moderate as $5\% \sim 10\%$.

The one loop SPT is the simplest among perturbation theories.
In principle there is no real obstacle in
adopting other theories advanced in recent years. As a lengthy but incomplete list, there are the 
renormalized perturbation theory 
\citep[e.g.][]{CrocceScoccimarro2006a, CrocceScoccimarro2006b, CrocceScoccimarro2008, BernardeauEtal2008}, 
the closure theory \citep{TaruyaHiramatsu2008, HiramatsuTaruya2009}, the renormalization group perturbation
theory \citep[e.g.][]{McDonald2007, MatarresePietroni2007, MatarresePietroni2008}, and many other variants to
these new techniques 
\citep[e.g.][]{Valageas2008, Pietroni2008, PietroniEtal2012, BernardeauEtal2012, CrocceEtal2012, AnselmiPietroni2012, TaruyaEtal2012, SugiyamaFutamase2012a, SugiyamaFutamase2012b}.  There are numerical codes
implementing some of these novel approaches made available to public, for example, 
the {\bf CLASS}\footnote{http://class-code.net} \citep{Lesgourgues2011}, 
the {\bf RegPT}\footnote{http://www-utap.phys.s.u-tokyo.ac.jp/\textasciitilde ataruya/regpt\_code.html} \citep{TaruyaEtal2012} and 
the {\bf MPTbreeze}\footnote{http://maia.ice.cat/crocce/MPTbreeze/} \citep{CrocceEtal2012}.
Development of momentum spectra in theories at SPT beyond 1-loop level is beyond scope of this paper,
but an intrinsic shortcoming of these perturbation theories is their ignorance of velocity vorticity, which
is likely the reason that these theories can not go deep into nonlinear regime. We notice that a recently developed 
semi-analytical theory, namely the effective field theory (EFT), could recover nonlinear evolution of
statistics beyond stream
crossing of the cosmic large scale structures much 
effectively \citep[e.g.][]{CarrascoEtal2012, BaldaufEtal2015, ForemanEtal2016}, which
is a practical solution to fulfill the demand on theory of the precision cosmology.

\begin{figure*}
\resizebox{\hsize}{!}{\includegraphics{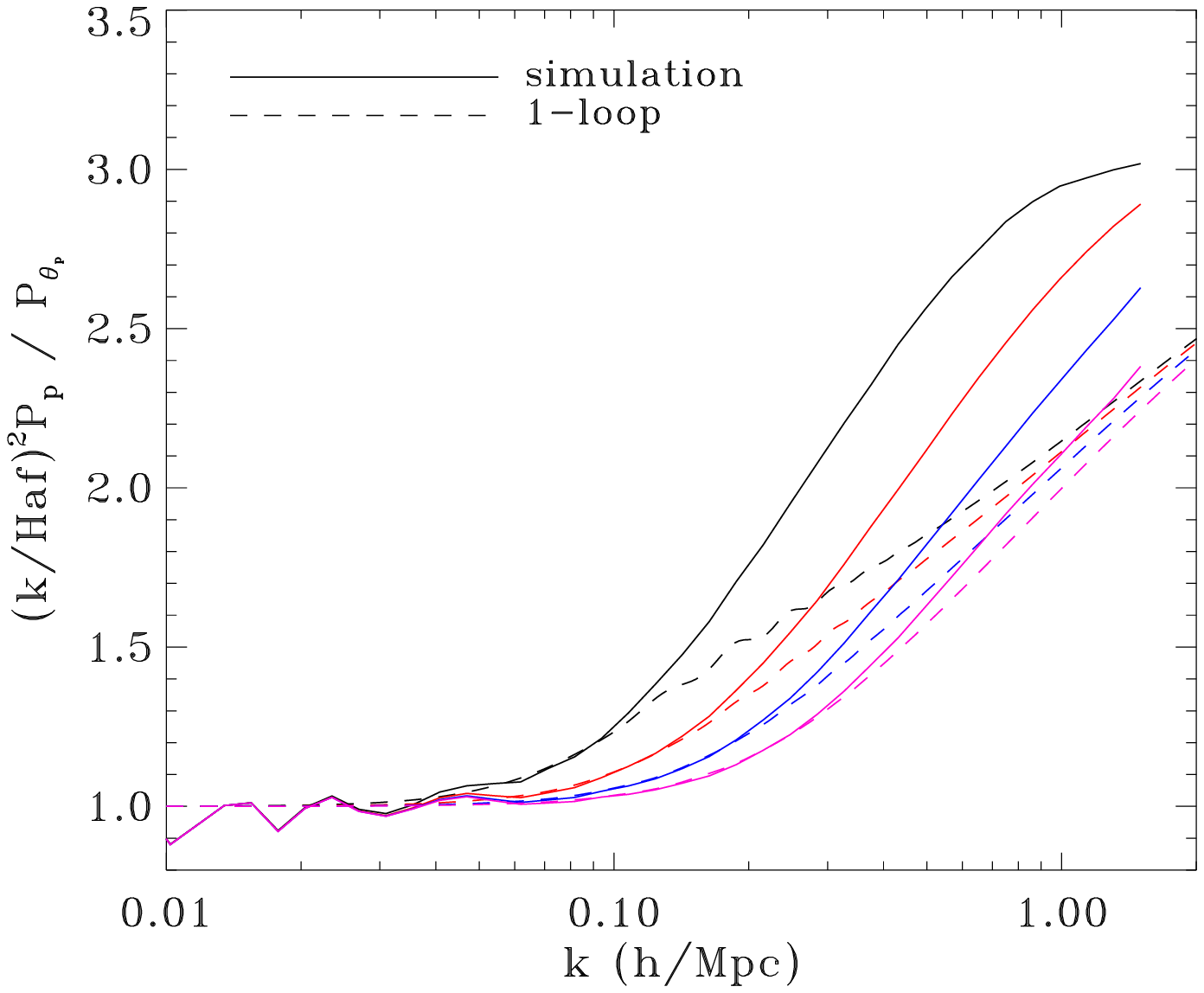}\includegraphics{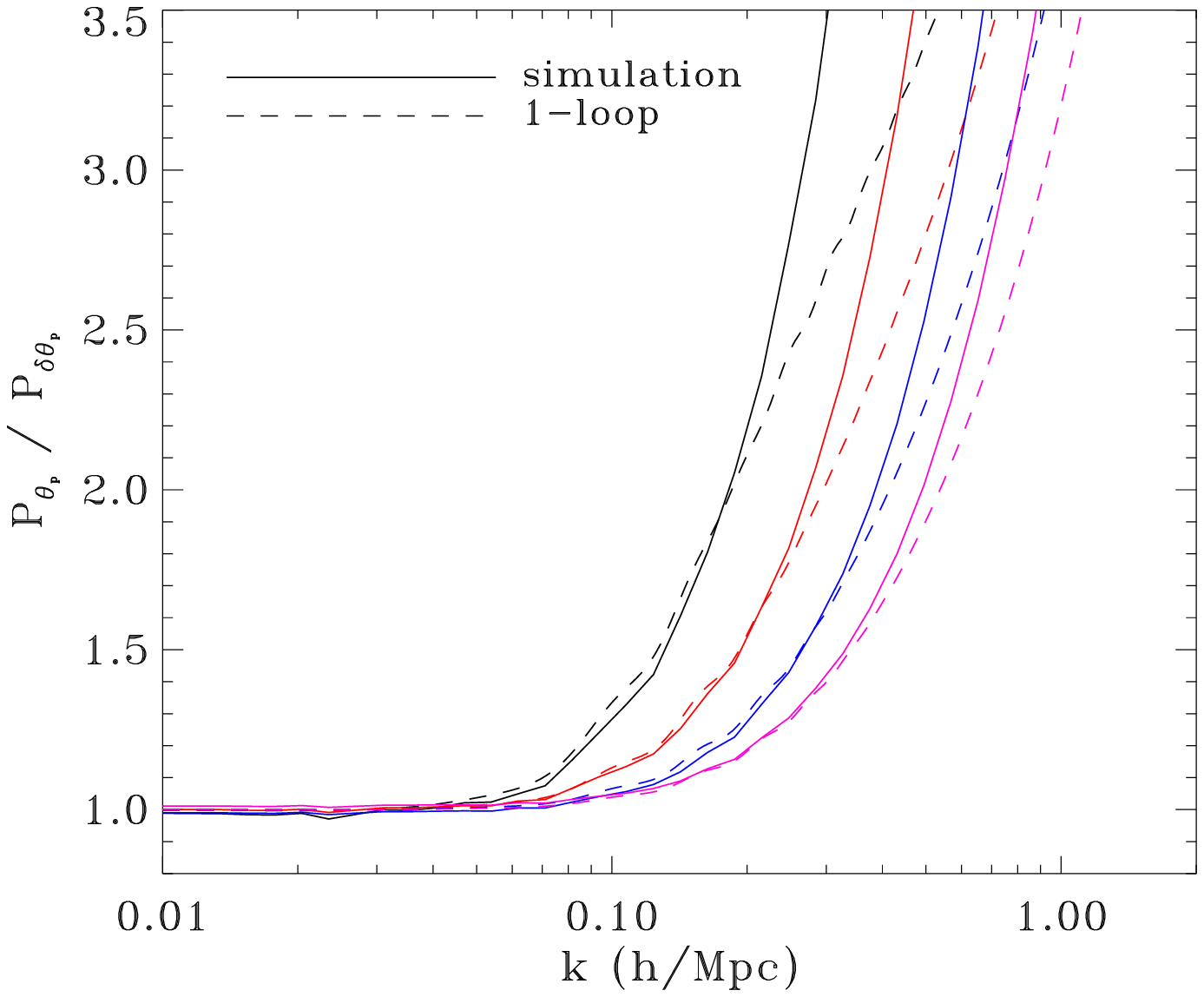}}
\caption{Differences between $P_p$, $P_{\theta_p}$ and $P_{\delta\theta_p}$
at four epochs of $z=0,1, 2, 3$ (in colors of black, red, blue and magenta 
correspondingly) respectively. Dashed lines are prediction of Eulerian
perturbation theory at one loop level (details in Appendix~\ref{sec:1loop}).}
\label{fig:ratios}
\end{figure*}

At large scales it is often assumed that the curl component of peculiar velocity field
is negligible, in principle one can reconstruct the vector velocity field from the its 
divergence field. But such operation is not applicable to the momentum field.
The vorticity of momentum contains component produced by the coupling between the 
spatial gradient of the density and the peculiar velocity,
\begin{equation}
\nabla \times \vect{p}=(1+\delta)\nabla \times \vect{v}+\nabla\delta \times \vect{v}\ .
\end{equation}
Obviously even if $\nabla \times \vect{v}=0$ as assumed generally in
perturbation theories, $\nabla \times \vect{p}\neq 0$, and the
$P_p$ is not equivalent to $P_{\theta_p}$ at all (left panel of Figure~\ref{fig:ratios}).
The rotational part in momentum in simulation indeed becomes very strong already in weakly nonlinear 
regime (Figure~\ref{fig:ratios}). Eulerian perturbation theory at one loop (Appendix~\ref{sec:1loop})
is invoked to check against simulation, the theory can only recover
$P_p/P_{\theta_p}$ at $z=0$ at scales $k< 0.1h/$Mpc. We can see that even in perturbation theory, 
the relation between the momentum and its divergence is complicated, actually we tried
several empirical proposals, but it seems there are no simple ways to recover $P_p$ from $P_{\theta_p}$.

Momentum divergence can be decomposed  as sum of two parts, namely a part $\theta_p^\delta$ 
which is fully correlated
with density fluctuation while the other one $\theta_p^S$ is not at all, $\theta_p=\theta_p^{\delta}+\theta_p^S$.
The cross correlation $P_{\delta\theta_p}$ is effectively $\avg{\delta\theta_p^\delta}$, which
does not contain any information of $\theta_p^S$. As
illustrated in the right panel of Figure~\ref{fig:ratios},
the power of $\theta_p^S$ is very large. We notice that 
$P_{\delta\theta_p}=\frac{1}{2f} \partial P_\delta/\partial \ln a$ is the 
time derivative of density power spectrum (Eq.~\ref{eq:Ppc}), thus it is viable to
straightforwardly derive the nonlinear $P_{\delta\theta_p}$ of dark matter 
from the nonlinear matter power spectrum
produced either by halo 
models \citep{CooraySheth2002, GiocoliEtal2010} 
or empirical formulas \citep{SmithEtal2003, TakahashiEtal2012, MeadEtal2015}. 
The bad news is that, as we attemped, there is no such simple scaling relation between $P_{\theta_p^S}$ and
$P_{\theta_p^\delta}$ as the one between $P_{\theta^\delta}$ and $P_{\theta^S}$ 
found in \citet{ZhengEtal2013}. One has to search for new ways to establish
link between nonlinear $P_{\theta_p}$ and $P_{\delta\theta_p}$.

%======================================================
\section{Discussion and conclusion}
In this report we present FFT based estimators for auto power spectra of momentum and
momentum divergence, and the cross spectrum of density fluctuation and momentum divergence. Although
these estimators are for ideal sample free of observational effects, which nevertheless can be readily 
incorporated to proposals handling with realistic observational samples.
Algorithms to clean alias effects using the third order Bettle-Lemari\'{e} scaling function are proposed and
thoroughly tested with simulation data sets, experiment
proves that the algorithm is able to preserve sub-percent precision till close to the Nyquist frequency. 

It is pointed out that non-zero bulk flow could induce additional shot noises, but that is 
only part of the story. Bulk flow might induce much more complicated effects as already
discussed in \citet{ParkPark2006} and \citet{Howlett2019}. Exact formulas are derived and numerically confirmed
to depicting the changes caused by removing bulk flow from peculiar velocities. 
Subtracting bulk flow results in generally minuscule changes to momentum power spectra at large scales, but might
has non-negligible significance in nonlinear regime, interestingly the real part of $P_{\delta\theta_p}$ is immune
to bulk flow. Numerical experiment suggests that there is no need to subtract bulk flow from
peculiar velocities for samples which peculiar velocities are exact or estimated with high accuracy.
However, we need to address that comprehensive treatment of impact of bulk flow on estimation of 
statistics of momentum is actually connected with the so called integral constraint problem, which is not considered
here, appropriate proposals to correct effects of bulk flow are left for future investigation.

To overcome the huge variances in power spectra at large scales due to limited
number of Fourier modes, momentum power spectra of the initial cosmic fields at $z=80$ of the simulation are
measured and linearly evolved to specified redshifts, which are then used as linear templates
to normalize measurements at those redshifts. The method greatly reduce the sample variances at large scales, making the comparison with theoretical models much
smooth and clear. Analysis of subsamples of our simulation shows that, cosmic variances of
$P_p$, $P_{\theta_p}$ and $P_{\delta\theta_p}$ are at $\sim 20\%$ level at large scales of $k<0.1h/$Mpc, being much larger than the cosmic variances of the density power spectrum. In nonlinear regime, cosmic variances of 
$P_p$ and $P_{\theta_p}$ keep at the same level, but the cosmic variances of $P_{\delta\theta_p}$ gradually 
decrease to $\sim10\%$ at $k>0.2h/$Mpc. 
A quick comparison of momentum power spectra of dark matter in simulation with theories indicates that if precision 
requirement is set to $\sim 10\%$, at large scales the one loop SPT agrees with simulation slightly
better than the linear theory at $z=0$. Of course, the performance of one loop SPT improves with increasing redshifts.

We also notice that $P_p$ contains strong power from the rotational part of momentum, and 
there is considerably large stochastic component in $\theta_p$ which is completely not
correlated with the density fluctuation. The two ingredients make it rather challenging to reconstruct
the full momentum field and its divergence beyond linear regime with the information offered by the density and
the cross correlation between density and momentum divergence, the three kinds of momentum
power spectra have their own distinctness.

\begin{acknowledgements}
JP is support by the National key R\&D program of China under 
grant no. 2018YFE0202900, and the NSFC grant of no. 11573030. It is greatly
appreciated that Dr. Li Ming realized the simulation used in this work, and Dr. Feng Longlong
kindly provided his code for reference. 
\end{acknowledgements}

%======================================================
\bibliographystyle{raa}

%======================================================
\appendix
\section{Prediction of Eulerian perturbation theory at one loop level on momentum power spectra}
\label{sec:1loop}
The momentum power spectrum in Fourier space can be expressed as
\begin{equation}
P_p=P_v+ \frac{1}{(2\pi)^3}\left[P\otimes P_v+P_{\delta v} \otimes P_{\delta v}^* \right]
+ 2\mathcal{B}_{\delta v v}+\mathcal{T}_{\delta v\delta v}\ ,
\label{eq:Ppdv}
\end{equation}
where $P_v(k)$ is the power spectrum of peculiar velocity, $P(k)$ is the matter power spectrum (sometimes denoted as $P_\delta$),  
$P_{\delta v}$ is the anisotropic cross-power spectrum of density contrast and peculiar velocity. In Eq.~\ref{eq:Ppdv} 
$\mathcal{B}_{\delta v v}$ and $\mathcal{T}_{\delta v \delta v}$ are integrations over bispectrum and trispectrum respectively
\begin{equation}
\begin{aligned}
\mathcal{B}_{\delta v v}&=
\frac{1}{(2\pi)^3}\int B_{\delta v v}(\vect{k}-\vect{q}, \vect{q}, -\vect{k}) \rmd \vect{q}\\
\mathcal{T}_{\delta v \delta v}
&=\frac{1}{(2\pi)^6}\int \int T_{\delta v \delta v}(\vect{k}-\vect{q}, \vect{q}, -\vect{k}-\vect{q}', \vect{q})\rmd \vect{q} \rmd \vect{q}'\ ,
\end{aligned}
\end{equation}
where $B_{\delta v v}(\vect{k}_1,\vect{k}_2,\vect{k}_3)\delta_D(\sum_i \vect{k}_i=0) \equiv \avg{\delta_1 \vect{v}_2 \cdot \vect{v}_3}_c$, 
$T_{\delta v \delta v}(\vect{k}_1,\vect{k}_2,\vect{k}_3,\vect{k}_4)\delta_D(\sum_i\vect{k}_i=0) \equiv \avg{\delta_1 \vect{v}_2 \cdot \delta_3 \vect{v}_4}_c$ 
, $\delta_D$ is the Dirac $\delta$-function, and $\avg{\ldots}_c$ refers to the irreducible correlation.

At scales $k\ll 1 h{\rm Mpc}^{-1}$ power spectrum of the curl 
component of velocity is an order of magnitude lower than the irrotational 
part \citep[e.g.][]{PueblasScoccimarro2009, ZhengEtal2013}, the velocity field 
can be approximated by the potential $\theta\equiv -\nabla \cdot \vect{v}/(Haf)$ alone.
In such ansatz there are
\begin{equation}
\begin{aligned}
&P_{\delta v}  =-i(Haf\vect{k}/k^2)P_{\delta\theta}\ , \ \ \ P_v=(Haf/k)^2 P_{\theta\theta}\ ,\\
&B_{\delta v v}(\vect{k}-\vect{q},\vect{q}, -\vect{k}) = 
(Haf)^2\frac{\vect{k}\cdot \vect{q}}{k^2 q^2}B_{\delta\theta\theta}\ ,\\
&T_{\delta v \delta v}(\vect{k}-\vect{q},\vect{q}, -\vect{k}-\vect{q}',\vect{q}') 
= -(Haf)^2 \frac{\vect{q}\cdot \vect{q}'}{q^2 q'^2}T_{\delta\theta\delta\theta}\ ,
\end{aligned}
\end{equation}
where $f\equiv \rmd \ln D(z)/\rmd \ln a$ with $D(z)$ being the linear density growth factor at redshift $z=1/a-1$.
The corresponding approximation to Eq.~\ref{eq:Ppdv} is then
\begin{equation}
\left( \frac{k}{Haf} \right)^2 P_p = 
P_{\theta\theta}  + \frac{k^2}{(2\pi)^3} 
\left[ P\otimes \left( \frac{P_{\theta\theta}}{k^2} \right)   + \left( \frac{\vect{k}P_{\delta\theta}}{k^2}\right) \otimes \left(\frac{\vect{k}P_{\delta\theta}}{k^2} \right)\right]
 +  
k^2\left( 2\mathcal{B}_{\delta\theta\theta}+\mathcal{T}_{\delta\theta\delta\theta}\right)\ ,
\label{eq:Ppd}
\end{equation}
where
\begin{equation}
\begin{aligned}
\mathcal{B}_{\delta\theta\theta}& =\frac{1}{(2\pi)^3}
\int \frac{\vect{k}\cdot \vect{q}}{k^2 q^2}
B_{\delta\theta\theta}(\vect{k}-\vect{q},\vect{q}, -\vect{k})\rmd \vect{q} \\
\mathcal{T}_{\delta\theta\delta\theta}&=\frac{1}{(2\pi)^6}\int\int
\frac{\vect{q}\cdot \vect{q}'}{q^2 q'^2} 
T_{\delta\theta\delta\theta}(\vect{k}-\vect{q},\vect{q},-\vect{k}-\vect{q}', \vect{q}')\rmd \vect{q} \rmd \vect{q}'\ .
\end{aligned}
\label{eq:bt}
\end{equation}
If $\delta$ and $\theta$ are both Gaussian, $\theta=\delta$,
$B_{\delta\theta\theta}=0$ and $T_{\delta\theta\delta\theta}=0$, Eq.~\ref{eq:Ppd} reduces to the
known Gaussian approximation \citep[e.g.][]{MaFry2002}, 
\begin{equation}
\left( \frac{k}{Haf}\right)^2 P_p^G=P_L+\frac{k^2}{(2\pi)^3}\left[P_L\otimes \left(\frac{P_L}{k^2}\right) + \left(\frac{\vect{k}P_L}{k^2}\right)
\otimes \left( \frac{\vect{k}P_L}{k^2}\right) \right]\ ,
\label{eq:PpdL}
\end{equation}
in which $P_L=D_z^2P_0$ with $P_0$ being the linear power spectrum at $z=0$, and $D_z=D(z)/D(z=0)$. 

At large scales where nonlinearity is weak one can invoke
perturbative theories, such as the standard Eulerian perturbation theory (SPT, Appendix~\ref{sec:appSPT}).
Implementing the SPT 
power spectra (Appendix~\ref{sec:appSPTP})  and
bispectrum (Appendix~\ref{sec:appSPTB}) to Eq.~\ref{eq:Ppd}, after truncation of terms of order higher than $D_z^4$,  
yields
\begin{equation}
\begin{aligned}
&\left( \frac{k}{Haf}\right)^2   P^{(1)}_p   =  D_z^2 P_0+ D_z^4 (P_{\theta\theta, 1}+P_{cov}+P_B) \\
&P_{cov}=P_{cov}^{\Rmnum{1}} +P_{cov}^{\Rmnum{2}}\ , \  \  P_B= 2k^2\mathcal{B}_{\delta\theta\theta,0} \\
&P_{cov}^{\Rmnum{1}} =\frac{k^2}{(2\pi)^3} P_0\otimes \left(\frac{P_0}{k^2}\right) \ , \ \
P_{cov}^{\Rmnum{2}} =\frac{k^2}{(2\pi)^3}  \left(\frac{\vect{k}P_0}{k^2}\right)\otimes \left( \frac{\vect{k}P_0}{k^2}\right)\ ,
\end{aligned}
\label{eq:PpdD4}
\end{equation}
in which $\mathcal{B}_{\delta\theta\theta,0}$ is given by Eq.~\ref{eq:intBktree}, explicit
formula to compute $P_{cov}$ is in Appendix~\ref{sec:appCOV}. 

The route leading to power spectrum of momentum divergence in SPT
is different. The starting point is the continuity equation
\begin{equation}
a \frac{\partial \delta(\vect{x},t)}{\partial t}+
\nabla \cdot \left\{   \left[ 1+\delta(\vect{x}, t) \right] \vect{v}(\vect{x},t)\right\} =0\ .
\label{eq:cont}
\end{equation}
\citet{Seljak1996} has already utilized the equation to derive the
power spectrum of the time derivative of the gravitational potential 
for investigation on Rees-Sciama effect. \citet{SmithEtal2009} also applied the same technique to measure 
integrated Sachs-Wolfe effect in N-body simulation. 
Fourier transforming Eq.~\ref{eq:cont} yields
$\theta_p(\vect{k})=\frac{1}{f}\partial \delta(\vect{k},a)/\partial \ln a$, corresponding power spectra are
\begin{equation}
\begin{aligned}
P_{\theta_p}(\vect{k})&=\frac{1}{f^2} 
\avg{ \frac{\partial \delta(\vect{k},a)}{\partial \ln a}\frac{\partial \delta^*(\vect{k},a)}{\partial \ln a}   } \\
P_{\delta\theta_p}(\vect{k})&=\frac{1}{f} 
\avg{\delta(\vect{k}) \frac{\partial \delta^*(\vect{k},a)}{\partial \ln a}}\ .
\end{aligned}
\label{eq:Ppc}
\end{equation}
Inserting the expansion scheme of SPT (Eq.~\ref{eq:SPTexp}),  $\delta=\sum_n D_z^n \delta_{(n)}$, 
there is 
\begin{equation}
\begin{aligned}
P_{\theta_p}  & =D_z^2P_0+\sum_{n} D_z^{2n}\left[  2 \sum_{j=1}^{n-1} j(2n-j) \langle \delta_{(j)} \delta^*_{(2n-j)} \rangle + n^2 \langle \delta_{(n)} \delta^*_{(n)} \rangle \right] \\
P_{\delta\theta_p}&=D_z^2P_0+\sum_{n} D_z^{2n} \left[  2 \sum_{j=1}^{n-1} n \langle \delta_{(j)} \delta^*_{(2n-j)} \rangle + n \langle \delta_{(n)} \delta^*_{(n)} \rangle \right] \ ,
\end{aligned}
\label{eq:Ppcexp}
\end{equation}
where we have used the property that odd order terms are zero.
The difference between $P_{\theta_p}$ or $P_{\delta\theta_p}$ and $P$ lies 
in coefficients associated with terms 
at different orders, it is very convenient to calculate momentum power spectrum: once higher
order correction terms to $P$ are ready, prediction fo $P_{\theta_p}$ or $P_{\delta\theta_p}$ can be
constructed simultaneously. $P_{\theta_p}$ and $P_{\delta\theta_p}$ to 
the order of $D_z^4$ are simply \citep{SmithEtal2009}
\begin{equation}
\begin{aligned}
P^{(1)}_{\theta_p}  & = D_z^2 P_0 +D_z^4 (6P_{13}+4P_{22}) \\
P^{(1)}_{\delta\theta_p} &= D_z^2 P_0 + D_z^4 (4P_{13}+2P_{22}) \ .
\end{aligned}
\label{eq:pdivpD4}
\end{equation}

%%====================================================
\section{Expansion scheme and computing formulas}
\label{sec:appSPT}
%%--------------------------------------------------------
\subsection{The expansion}
In SPT, $\delta_k$ and $\theta_k$ are expanded as
\begin{equation}
\delta_k=\sum_{n=1}^\infty D_z^n \delta_{(n)}\ , \ \ \ \theta_k=\sum_{n=1}^\infty D_z^n \theta_{(n)}\ ,
\label{eq:SPTexp}
\end{equation}
$\delta_{(1)}=\theta_{(1)}$ are simply linear quantities, higher order terms are constructed via
\begin{equation}
\begin{aligned}
\delta_{(n)}&=\frac{1}{(2\pi)^n}\int d^3 q_1 \ldots d^3 q_n 
\delta_D(\sum_i \vect{q}_i -\vect{k})F_n
\delta_{(1)}(\vect{q}_1)\ldots \delta_{(1)}(\vect{q}_n)       \\
\theta_{(n)}&=\frac{1}{(2\pi)^n}\int d^3 q_1 \ldots d^3 q_n 
\delta_D(\sum_i \vect{q}_i -\vect{k})G_n
\delta_{(1)}(\vect{q}_1)\ldots \delta_{(1)}(\vect{q}_n) \ .
\end{aligned}
\end{equation}
The kernels $F_n$ and $G_n$ are homogeneous functions of wave 
vectors $\{\vect{q}_1,\ldots,\vect{q}_n\}$, of which explicit formulas can 
be found in \citet{GoroffEtal1986} and \citet{JainBertschinger1994}.
 
%%------------------------------------------------------------
\subsection{Power spectra of $\delta$ and $\theta$}
\label{sec:appSPTP}
Power spectra in the framework are organized in the form of
\begin{equation}
P_{xy}=\sum_{n=1}^\infty D_z^{2n} P_{xy,n-1}\ , \ \ \ 
P_{xy,n-1} \equiv 2\sum_{j=1}^{n-1} \avg{x_{(j)} y^*_{(2n-j)}} +\avg{x_{(n)} y_{(n)}^*}\ ,
\label{eq:SPTPk}
\end{equation}
where $x,y$ represent $\delta$ or $\theta$ (for $\delta$--$\delta$ subscript is omitted by default in this paper), 
$P_{xy,0}=P_0$ is the linear matter power spectrum.
Power spectrum corrected to 2-loop level is thus
\begin{equation}
\begin{aligned}
P_{xy}^{(2)} & =D_z^2 P_0+D_z^4 P_{xy,1}+D_z^6 P_{xy,2} = P_{xy}^{(1)}+D_z^6 P_{xy,2} \\
P_{xy,1} & =2P_{xy,13}+P_{xy,22} \\
P_{xy,2} & = 2P_{xy,15}+2P_{xy,24}+P_{xy,33}\ ,
\end{aligned}
\label{eq:Pk2loop}
\end{equation}
in which explicit expressions of $P_{xy, ij}=\avg{x_{(i)}y^*_{(j)}}$ can be found in e.g.
\citet{BernardeauEtal2002}, \citet{CarlsonEtal2009} and \citet{TaruyaEtal2009}. Then 1-loop corrections to
power spectra are
\begin{equation}
\begin{aligned}
P_{13} & = \frac{P_0(k)}{504}  \frac{k^3}{4\pi^2} \int_0^\infty \rmd r P_0(kr)\left[ \frac{12}{r^2}-158
+100 r^2 -42r^4 + \frac{3}{r^2} (r^2-1)^3 (7 r^2+2)\ln \abs{\frac{1+r}{1-r}} \right] \\
P_{\delta\theta,13} & = \frac{P_0(k)}{504} \frac{k^3}{4\pi^2}  \int_0^\infty \rmd r P_0(kr)
\left[ \frac{24}{r^2}-202+56r^2-30r^4 + \frac{3}{r^2}(r^2-1)^3 (5r^2+4) \ln \abs{\frac{1+r}{1-r}} \right]
\\
P_{\theta\theta,13}  & = \frac{P_0(k)}{168}  \frac{k^3}{4\pi^2}  \int_0^\infty \rmd r P_0(kr)
\left[ \frac{12}{r^2} -82 +4 r^2 -6r^4 + \frac{3}{r^2}(r^2-1)^3 (r^2+2) \ln \abs{\frac{1+r}{1-r}} \right]
\end{aligned}
\end{equation}
and
\begin{equation}
\begin{aligned}
P_{22} &= \frac{1}{98} \frac{k^3}{4\pi^2}  \int_0^\infty \rmd r P_0(kr) \int_{-1}^1 \rmd x
P_0\left( k\sqrt{1+r^2-2rx} \right) \cdot \frac{(3r+7x-10rx^2)^2}{(1+r^2-2rx)^2} \\
P_{\delta\theta, 22} & = \frac{1}{98} \frac{k^3}{4\pi^2}  \int_0^\infty \rmd r P_0(kr) \int_{-1}^1 \rmd x
P_0\left( k\sqrt{1+r^2-2rx} \right) \cdot \frac{(3r+7x-10rx^2)(7x-r-6rx^2)}{(1+r^2-2rx)^2} \\
P_{\theta\theta, 22} & = \frac{1}{98} \frac{k^3}{4\pi^2}  \int_0^\infty \rmd r P_0(kr) \int_{-1}^1 \rmd x
P_0\left( k\sqrt{1+r^2-2rx} \right) \cdot \frac{(7x-r-6rx^2)^2}{(1+r^2-2rx)^2} \ .
\end{aligned}
\end{equation}

%%--------------------------------------------------------
\subsection{The density-velocity-velocity bispectrum}
\label{sec:appSPTB}
The loop expansion for the density-velocity-velocity bispectrum can be written down following
\citet{Scoccimarro1997} and \citet{ScoccimarroEtal1998},
\begin{equation}
B_{\delta\theta\theta}=D_z^4 B_{\delta\theta\theta,0}+D_z^6 B_{\delta\theta\theta,1}+\ldots\ ,
\label{eq:Bk1loop}
\end{equation}
in which the tree-level bispectrum is
\begin{equation}
\begin{aligned}
B_{\delta\theta\theta,0}(\vect{k}_1,\vect{k}_2,\vect{k}_3)= 2P_0(k_1)P_0(k_2)G_2(\vect{k}_1,\vect{k}_2)+&2P_0(k_1)P_0(k_3)G_2(\vect{k}_1,\vect{k}_3) \\
&+2P_0(k_2)P_0(k_3)F_2(\vect{k}_2,\vect{k}_3)\ .
\end{aligned}
\end{equation}
$\mathcal{B}_{\delta\theta\theta}$ as the integral of the bispectrum $B_{\delta\theta\theta}$ is therefore
$\mathcal{B}_{\delta\theta\theta}=D_z^4\mathcal{B}_{\delta\theta\theta,0}+D_z^6\mathcal{B}_{\delta\theta\theta,1}+\ldots$, and 
\begin{equation}
\mathcal{B}_{\delta\theta\theta,0}=I_1+I_2+I_3\ ,
\label{eq:intBktree}
\end{equation}
where
\begin{equation}
\begin{aligned}
I_1&=-\frac{1}{7}\frac{k}{4\pi^2}\int_0^\infty \rmd r P_0(kr)\int_{-1}^1 \rmd x P_0\left( k \sqrt{1+r^2-2rx}\right)
\frac{x(r-7x+6r x^2)}{1+r^2-2rx} \\
I_2&=\frac{1}{7}\frac{k}{4\pi^2}P_0(k)\int_0^\infty \rmd r\int_{-1}^1\rmd x P_0\left( k \sqrt{1+r^2-2rx}\right) 
\frac{r^3x(7rx-1-6x^2)}{1+r^2-2rx}\\
I_3&=-\frac{2}{3}\frac{k}{4\pi^2}P_0(k)\int_0^\infty ( r^2+1)P_0(kr) \rmd r\ .
\end{aligned}
\end{equation}

%---------------------------------------------------------------------------------
\subsection{Convolution terms}
\label{sec:appCOV}

\begin{equation}
\begin{aligned}
P_{cov}^{\Rmnum{1}} & =\frac{k^3}{4\pi^2}   \int_0^\infty  \rmd r P_0(kr) 
\int_{-1}^1 P_0\left( k\sqrt{1+r^2-2rx} \right)  \rmd x \\
P_{cov}^{\Rmnum{2}} &= 
 \frac{k^3}{4\pi^2}\int_0^\infty \rmd r P_0(kr) \int_{-1}^1  P_0\left(k\sqrt{1+r^2-2rx}\right) \frac{r(x-r)}{1+r^2-2rx} \rmd x \ .
\end{aligned}
\label{eq:Pcov}
\end{equation}

\end{document}